\documentclass[nofootinbib,prx,twocolumn,showpacs,superscriptaddress,notitlepage,superscriptaddress,amsmath]{revtex4-2}
\usepackage[T1]{fontenc}

\usepackage{lipsum}  
\usepackage{dsfont}
\usepackage{amsmath}
\usepackage{braket}
\usepackage{amssymb}
\usepackage{amsthm}
\usepackage{algpseudocode}
\usepackage{algorithm}
\usepackage{siunitx}
\usepackage{amsfonts}
\usepackage{comment}
\usepackage[normalem]{ulem}
\usepackage{graphicx}
\usepackage{color,framed}
\usepackage{hyperref}
\usepackage{enumerate}
\usepackage{slashed}
\usepackage{url}
\usepackage{bbm}
\usepackage{chngcntr}
\counterwithout{equation}{section}
\usepackage{tikz,pgfplots}
\usepackage{pgfplotstable}
\usepackage{siunitx}
\usepackage{comment}
\usepackage{graphicx}

\makeatletter
\newcommand{\setlabel}[1]{\edef\@currentlabel{#1}\label}
\makeatother

\usepgfplotslibrary{fillbetween}
\hypersetup{
    colorlinks=true,
    linkcolor=blue,
    filecolor=magenta,      
    urlcolor=cyan,
}

\newcommand{\sinc}{\,{\rm sinc}}

\hypersetup{
    colorlinks=true, 
    linktoc=all,     
    linkcolor=blue,  
}
\def \beq {\begin{equation}}
\def \eeq {\end{equation}}
\def \beqa {\begin{eqnarray}}
\def \eeqa {\end{eqnarray}}
\def \bseq {\begin{subequations}}
\def \eseq {\end{subequations}}

\newcommand \tr {\,{\rm tr}}

\newcommand{\argtanh}{{\rm arctanh}\,}

\newcommand{\D}[1]{\text{d}#1}

\pgfplotsset{compat=1.18}
\begin{document}

\title{Adiabatic preparation of thermal states \\
and entropy-noise relation on noisy quantum computers}

\author{Etienne Granet}
\email{etienne.granet@quantinuum.com}
\author{Henrik Dreyer}
\affiliation{Quantinuum, Leopoldstrasse 180, 80804 Munich, Germany}
\date{\today}

\begin{abstract}

We consider the problem of preparing thermal equilibrium states at finite temperature on quantum computers. Assuming thermalization, we show that states that are \emph{locally} at thermal equilibrium can be prepared by evolving adiabatically an initial  thermal Gibbs state of a simple Hamiltonian with an interpolating time-dependent Hamiltonian, identically to adiabatic ground state preparation. We argue that the entropy density of \emph{local} density matrices is conserved during the adiabatic evolution in the thermodynamic limit, so that both the entropy and energy of the final state can be computed, and thus the final temperature too. We show that in the presence of hardware noise, the entropy created by the noisy evolution can be precisely benchmarked with mirror circuits. We give numerical evidence that the resulting thermal state preparation protocol is \emph{noise-resilient} for depolarizing noise, in the sense that the energy-temperature curve measured on a noisy quantum computer is remarkably insensitive to the amplitude of depolarizing noise in the state preparation. We finally propose a protocol to estimate the lack of adiabaticity in a given actual Trotter implementation of the dynamics. We test our protocol on Quantinuum's H1-1 ion-trap device. We measure that a circuit with $640$ two-qubit gates implemented on hardware generates an entropy per site of $0.166 \pm 0.0045$, giving a benchmark metric for this state preparation. We report the preparation of a thermal state with temperature $2.56  \pm 0.26$ of the Ising model in size $5\times 4$.

\end{abstract}

\maketitle

\section{Introduction} 
The simulation of materials and condensed matter systems is expected to be one of the first applications of quantum computers \cite{dalzell2023quantum,feynman2018simulating}. While low-entanglement ground state physics can be studied classically in a number of cases albeit with extensive numerical effort \cite{ponsioen2019period,qin2020absence,xu2024coexistence}, any settings that involve higher-excited states is considerably more challenging to classical computers. This includes computing the out-of-equilibrium dynamics of a system, for which state-of-the-art hardware is already at the frontier of the classically simulable regime, or making progress towards it \cite{haghshenas2025digital,abanin2025constructive,andersen2025thermalization,king2025beyond,kim2023evidence,granet2025superconducting}. But this also includes \emph{finite-temperature} equilibrium properties, whenever the system cannot be studied with quantum Monte Carlo techniques \cite{foulkes2001quantum}. 

While there exist multiple standard quantum algorithms for preparing the ground state of a Hamiltonian (such as the quantum adiabatic algorithm \cite{albash2018adiabatic} or quantum phase estimation \cite{kitaev1995quantum}), fewer standard algorithms are known for the preparation of finite-temperature states. There exist imaginary time evolution algorithms for preparing Gibbs states \cite{chowdhury2016quantum,motta2020determining,coopmans2023predicting,shtanko2021preparing}, sampling-based techniques \cite{temme2011quantum,cohn2020minimal}, Lindbladian evolution techniques or interaction with an external bath \cite{chen2023efficient,shtanko2021preparing,granet2024noise,bergamaschi2024quantum,chen2023quantum,brunner2024lindblad,rall2023thermal,haug2022generalized}, variational quantum algorithms or optimized parametrized circuits \cite{consiglio2024variational,sewell2022thermal,martyn2019product,wu2019variational,chowdhury2020variational,wang2021variational,warren2022adaptive,foldager2022noise}, filtering methods \cite{lu2021algorithms}, techniques with linear combination of unitaries \cite{holmes2022quantum,chowdhury2016quantum}, or using thermalization \cite{poulin2009sampling,terhal2000problem,riera2012thermalization,bilgin2010preparing}. It is however not clear how practical some of these approaches are, since they require for example some efficient low-energy state preparation, a large number of measurements or a large number of sampling over random states, see e.g. \cite{hemery2024measuring} for a hardware implementation of one of them.

In this paper, we study the generalization of the adiabatic state preparation of ground states to finite-temperature states. We relax the objective of preparing thermal Gibbs states $e^{-\beta H}$, to only preparing states that are \emph{locally} at equilibrium. This relaxed condition is justified in two ways. Firstly, on physical grounds, for a closed quantum system, only \emph{local} thermal equilibrium can be obtained by purely unitary evolution. Secondly, the exact preparation of the Gibbs state on the entire system might take a very long time, due to the exponentially large number of exponentially close energy levels in the spectrum. We explain that we can prepare states that are locally at thermal equilibrium by applying an adiabatic evolution between a simple Hamiltonian and the target Hamiltonian, starting from a thermal state of the initial simple Hamiltonian. Our protocol crucially relies on thermalization \cite{mori2018thermalization}. We then show how to compute the temperature of the final state produced, using the fact that an ideal adiabatic evolution preserves the entropy density of local reduced density matrices in the thermodynamic limit, for which we give a derivation.

Actual quantum computing hardware always contains imperfections that limit the number of gates that can be implemented. This noise has the effect of increasing the entropy of the density matrix describing the state of the quantum computer. In particular, the adiabatic evolution implemented cannot be isentropic in the presence of noise. We propose a method to benchmark the increase in entropy of the state during the noisy time evolution. With this method, we show with numerical simulations with depolarizing noise that our thermal state preparation is \emph{noise-resilient}, in the sense that a state prepared on a noisy hardware at a given temperature is equivalent to a noiseless state preparation at a higher temperature. However, noise constrains the minimal temperature that can be reached by the protocol. We also propose a method to estimate the lack of adiabaticity in the evolution. Finally, we test the protocol that we describe on Quantinuum's H1-1 ion-trap device, measure the entropy injected by hardware noise and compute the temperature of the state prepared.

Applying the adiabatic algorithm beyond the preparation of ground states has been studied in a number of works before. The preparation of excited states has been studied in \cite{yarloo2024adiabatic}. The preparation of thermal Gibbs states has been considered in \cite{greenblatt2024adiabatic} at low temperatures and in \cite{il2021adiabatic,zuo2024work,irmejs2025quasi} for general temperatures. Special cases have been studied as well \cite{plastina2014irreversible,francica2019role}. These different works have however focused on preparing the exact Gibbs states on the entire system. The main novelties of this work are (i) a protocol to prepare states that are only \emph{locally} thermal, using thermalization arguments and the fact that the ideal adiabatic evolution is locally isentropic, (ii) the study of the effect of noise on the entropy density of the system, and the finding that the energy/temperature curves obtained are robust to noise, and (iii) the implementation of this thermal adiabatic evolution on actual quantum computing hardware.

\section{Adiabatic evolution of thermal states} 
\subsection{Definitions} 
Given a Hamiltonian $H$ on $N$ sites and an inverse temperature $\beta$, the free energy of a density matrix $\rho$ is defined as
\begin{equation}\label{freeenergy}
    \mathcal{F}[\rho]=\mathcal{E}[\rho]-\frac{1}{\beta }\mathcal{S}[\rho]\,,
\end{equation}
where $\mathcal{E}[\rho]=\tr[\rho H]$ is the energy, and where the entropy $\mathcal{S}[\rho]$ of the density matrix $\rho$ is
\begin{equation}
    \mathcal{S}[\rho]=-\tr[\rho \log \rho]\,.
\end{equation}
It is well known that the density matrix $\rho$ that minimizes the free energy $\mathcal{F}[\rho]$ for a given energy $\mathcal{E}[\rho]$ is the Gibbs state \cite{mori2018thermalization}
\begin{equation}\label{gibbs}
    \rho=\frac{e^{-\beta H}}{\tr[e^{-\beta H}]}\,.
\end{equation}
It can be interpreted equivalently as the state that maximizes the entropy $\mathcal{S}$ at fixed energy $\mathcal{E}$. According to general statistical physics ideas, entropy is expected to get maximized when systems freely evolve, up to the constraints imposed by the conserved quantities of their Hamiltonian. The free energy $\mathcal{F}[\rho]$ as defined in \eqref{freeenergy} for the state $\rho$ on the entire system however does not satisfy this property, and is instead constant $\mathcal{F}[e^{iHt}\rho e^{-iHt}]=\mathcal{F}[\rho]$ for any $t$, because under any unitary evolution $U$ the entropy is conserved $\mathcal{S}[U\rho U^\dagger]=\mathcal{S}[\rho]$, and under evolution by the Hamiltonian $H$ the energy is conserved $\tr[H e^{iHt}\rho e^{-iHt}]=\tr[H\rho]$. This suggests that requiring an algorithm to prepare a Gibbs state on the entire system is not particularly physical, in the sense that there is already no ``natural'' way for a state with same initial energy to get there. Instead, a more physical objective is to only require that $\rho$ thermalizes \emph{locally}. This means that subsystems that are sufficiently large, but still small compared to the entire system, would be thermal states when the rest of the system is traced out, like if it was an environment \cite{goldstein2006canonical,popescu2006entanglement}. Given a subset $A$ of sites, we denote $O_A$ an observable only supported on the sites of $A$. We say that a Hamiltonian satisfies the thermalization property if for every density matrix $\rho$ with low energy variance, the time average of observables on $A$ converges at late times to that of a Gibbs state
\begin{equation}
    \frac{1}{T}\int_0^T \tr[O_Ae^{iHt}\rho e^{-iHt}] \D{t}\underset{T\to\infty}{\longrightarrow} \frac{\tr[O_A e^{-\beta H}]}{\tr[e^{-\beta H}]}\,,
\end{equation}
provided that the number $N_A$ of sites in the subset $A$ satisfies $1\ll N_A \ll N$. Here, $\beta$ is a parameter that depends on the initial density matrix. We have to impose that the variance of the energy $\tr[H^2 \rho]-\tr[H \rho]^2$ scales strictly smaller than $\mathcal{O}(N^2)$, otherwise the system would contain different macroscopic states with different energy densities. We note that the quantity $\tr[O_Ae^{iHt}\rho e^{-iHt}]$ is equal to $\tr_A[O_A \rho_A(t)]$, where we defined $\tr_A$ as the trace over the sites in $A$, and the local density matrix
\begin{equation}
    \rho_A(t)=\tr_{\bar{A}}[e^{iHt}\rho e^{-iHt}]\,,
\end{equation}
with $\tr_{\bar{A}}$ being the trace over the sites not in $A$. We will call $\rho$ a thermal state on $A$ if it satisfies
\begin{equation}
   \tr[\rho O_A]= \frac{\tr[O_A e^{-\beta H}]}{\tr[e^{-\beta H}]}\,,
\end{equation}
for any observable $O_A$ on $A$, in the limit $1 \ll N_A \ll N$. The unitary evolution on $\rho$ induces a non-unitary evolution on $\rho_A$, and the free energy and the entropy of $\rho_A$ are in general not constant under time evolution. When the initial state is a pure state, $\mathcal{S}[\rho_A]$ is called entanglement entropy of $\rho_A$ with the rest of the system. However, when the initial state is itself a mixed state, contributions to $\mathcal{S}[\rho_A]$ come both from this entanglement entropy and from the usual 
thermodynamic entropy. As the objective of the paper is to introduce a method to prepare states that are \emph{locally} at equilibrium, we will mostly deal with the entropy of the \emph{reduced} density matrix $\mathcal{S}[\rho_A]$.

We introduce specific notations to denote the energy and entropy density of a state. The energy density $E[\rho]$ is defined as
\begin{equation}
    E[\rho]=\frac{1}{N}\tr[H \rho]\,.
\end{equation}
What Hamiltonian $H$ is denoted here will be clear or specified in the context. The entropy density $S[\rho]$ is defined as
\begin{equation}\label{sdensity}
    S[\rho]=-\frac{1}{N_A}\tr_A[\rho_A \log \rho_A]\,,
\end{equation}
for an arbitrary subset of sites $A$ such that $N_A\gg 1$ and $N-N_A\gg 1$. The notation assumes here that the right-hand side is independent of $A$ for generic large simple subsets $A$ in the thermodynamic limit, which is a property that is expected for equilibrium states of uniform Hamiltonians. In all the rest of the paper, we will partition the system into two halves $A=\{1,...,N/2\}$.

\subsection{Thermal states remain local thermal states under adiabatic evolution} 
We are now going to argue that thermal states remain local thermal states under adiabatic evolution. We consider $H_0$ and $H_f$ two local Hamiltonians defined on $N$ qubits, as well as a time-dependent local Hamiltonian $H(t)$ that interpolates smoothly between $H(t=0)=H_0$ and $H(t=1)=H_f$. To fix the ideas and without loss of generality, we are going to consider the scheduling
\begin{equation}\label{pathadiabatic}
    H(t)=(1-t)H_0+tH_f\,,
\end{equation}
with the initial Hamiltonian
\begin{equation}\label{h0}
        H_0=-\sum_{j=1}^N X_j\,.
    \end{equation}
Any other time scheduling and simple initial Hamiltonian could be considered as well. We assume that for all $0\leq t\leq 1$, $H(t)$ and $-H(t)$ are gapped (at fixed value of system size), and that for $0<t<1$ the Hamiltonian $H(t)$ satisfies the thermalization property. We consider an initial density matrix that is a thermal state for the initial Hamiltonian $H_0$ at inverse temperature $\beta_0$. These Gibbs states $\rho_0\propto e^{-\beta_0 H_0}$ are product states in the $X$ basis
\begin{equation}
    \begin{aligned}
        \rho_0&=\bigotimes_{j=1}^N \left(\frac{I}{2}+\frac{\tanh\beta_0}{2}X_j \right)\,.
    \end{aligned}
    \end{equation}
    They can be prepared efficiently on a quantum computer as a statistical random mixture of pure product states, generated by initializing $|\pm\rangle_j$ for every site $j$ with probability $\frac{1\pm \tanh \beta_0}{2}$ for every shot run on the quantum computer. Then, for a given $T$ called adiabatic time, we evolve $\rho$ starting from $\rho(t=0)=\rho_0$ according to
\begin{equation}
    i\partial_s \rho=[H(s/T),\rho]\,,\qquad 0\leq s\leq T\,.
\end{equation}
We define the final density matrix $\rho_f=\rho(s=T)$. It is a function of $T$ itself.

Our result is that when $T\to\infty$, $\rho_f(T)$ is locally a thermal state for a certain inverse temperature $\beta_f$. The final inverse temperature $\beta_f$ is a function of the initial inverse temperature $\beta_0$, and for every $\beta_f$ there exists a $\beta_0$ such that the state evolves to a Gibbs state with inverse temperature $\beta_f$. This function $\beta_f(\beta_0)$ is given by
\begin{equation}\label{beta1}
    \beta_f=\frac{\partial_{\beta_0}S_0}{\partial_{\beta_0}E_f}\,,
\end{equation}
    where $E_f$ is the energy density of the final state with respect to $H_f$, i.e. $E_f=\frac{1}{N}\tr[\rho_f H_f]$, and $S_0$ is the entropy density of the \emph{initial} state $S_0=S[\rho_0]$. In particular, for $H_0$ given in \eqref{h0}, we have 
    \begin{equation}\label{finalbeta}
        \beta_f=-\frac{\beta_0 (1-\tanh^2 \beta_0)}{\partial_{\beta_0}E_f}\,.
    \end{equation}
    We note that given a final temperature $\beta_f$, one does not know \emph{a priori} what initial temperature $\beta_0$ is required to prepare it. However, as the energy/temperature curve of most Hamiltonians is monotonous, one can envision to target a specific final temperature with just a few repetitions of the protocol.\\

    These results are justified as follows. In the very slow adiabatic limit $T\to\infty$, along the path $H(t)$, the density matrix at time $t$ is evolved for an increasingly long time with a Hamiltonian close to $H(t)$. Since $H(t)$ satisfies the thermalization property by assumption, $\rho(t)$ is locally in a thermal state for $H(t)$ at all times. Hence it is in a thermal state for $H_f$ at some inverse temperature $\beta_f$ at the end of the path. Let us denote $U$ the unitary operator that implements this adiabatic evolution in the limit $T\to\infty$. The energy density of the final state is
    \begin{equation}
        E_f(\beta_0)=\frac{1}{N}\tr \left[H_f U\frac{e^{-\beta_0 H_0}}{\tr[e^{-\beta_0 H_0}]}U^\dagger \right]\,.
    \end{equation}
    This is a smooth function of $\beta_0$. Moreover, in the limit $\beta_0\to\infty$, $\rho_0$ is in the ground state of $H_0$, so according to the adiabatic theorem, $\rho_f$ is in the ground state of $H_f$ because there is always a gap along the path by assumption. Hence $\beta_f\to \infty$ when $\beta_0\to\infty$. Similarly, $\beta_f\to-\infty$ when $\beta_0\to-\infty$. We conclude that for any $\beta_f$, there is always a $\beta_0$ such that the final inverse temperature is $\beta_f$. Namely, it is always possible to prepare a state that is locally thermal for the final Hamiltonian $H_f$ with any inverse temperature $-\infty<\beta_f<\infty$, provided the adiabatic evolution starts from a thermal state of $H_0$ with a well-chosen inverse temperature $\beta_0$. This $\beta_f$ is a priori expressed as
    \begin{equation}
        \beta_f=\frac{\partial_{\beta_0}S_f}{\partial_{\beta_0}E_f}\,,
    \end{equation}
where $S_f=S[\rho_f]$ is the entropy density of the \emph{final} state. As we are only requiring to prepare states that are \emph{locally} thermal states, the entropy density is computed here for local reduced density matrices, as defined in \eqref{sdensity}. The entropy density is thus not necessarily conserved by unitary evolution, contrary to the entropy of the entire system. To get our result \eqref{beta1}, we now must show that the entropy density of the state remains constant along the adiabatic evolution in the thermodynamic limit.

\subsection{Adiabatic evolution is isentropic} 

In this section we show that the entropy density of the reduced density matrix remains constant during the adiabatic evolution of the thermal state in the thermodynamic limit. As stated before, this is a different affirmation than the entropy of the entire density matrix being constant -- which is guaranteed by the unitarity of the evolution. The entropy density of local reduced density matrices is generically not constant under time evolution \cite{calabrese2005evolution}. Although the fact that an ideal quantum adiabatic evolution preserves the  entropy of reduced density matrices is frequently mentioned by experimentalists, see e.g. \cite{spar2022realization,carcy2021certifying,bloch2008many}, this fact seems to be very rarely mentioned in the theory literature, to the best of our knowledge. We thus find it relevant to write down a derivation.

We fix a subset of sites $A$ with $N_A$ sites, and denote $\partial A$ the boundary of $A$, namely the sites in $A$ where there is a term in the Hamiltonian $H$ that touches both a site in $A$ and a site outside of $A$, and $|\partial A|$ the number of sites in $\partial A$. We are going to assume that the bulk of $A$ is much bigger than its boundary, namely that $|\partial A|\ll N_A$. Let us consider $\rho$ a Gibbs state $\rho=\frac{e^{-\beta H}}{\tr[e^{-\beta H}]}$. If the number of sites in $A$ is large enough, and if $H$ is local, we can assume that the reduced density matrix $\rho_A$ is also a Gibbs state with same temperature
\begin{equation}
    \rho_A=\frac{e^{-\beta H_A}}{\tr_A[e^{-\beta H_A}]}\,,
\end{equation}
with $H_A$ the Hamiltonian $H$ restricted to terms that are entirely in the region $A$, for $N_A\gg 1$. We now evolve $\rho$ for a time $t$ under $H$ perturbed by $\delta H$. The time evolution operator $e^{it(H+\delta H)}$ is, at first order in $\delta H$
\begin{equation}\label{correction}
    e^{it(H+\delta H)}=e^{itH}+\int_0^t e^{isH}\delta H e^{i(t-s)H}\D{s}+\mathcal{O}(\delta H^2)\,.
\end{equation}
We thus get
\begin{equation}
\begin{aligned}
    &\rho(t)=e^{it(H+\delta H)}\rho e^{-it(H+\delta H)}\\
    &=e^{itH}\rho e^{-itH}+\int_0^t e^{isH}[\delta H,e^{i(t-s) H}\rho e^{-i(t-s)H}] e^{-isH}\D{s}\\
    &+\mathcal{O}(\delta H^2)\,.
\end{aligned}
\end{equation}
Using that $H$ commutes with $\rho$, this is
\begin{equation}
    \rho(t)=\rho+\int_0^t e^{isH}i[\delta H,\rho ] e^{-isH}\D{s}+\mathcal{O}(\delta H^2)\,.
\end{equation}
This perturbation to $\rho(t)$ is traceless, because the trace of the total density matrix is conserved under unitary operation. For a small traceless perturbation $\delta \rho$, the entropy density on the sub-system $A$ is perturbed as
\begin{equation}
    S[\rho+\delta \rho]=S[\rho]-\frac{1}{N_A}\tr_A(\delta \rho_A \log \rho_A)\,.
\end{equation}
In our case, using that $\tr_A(\delta \rho_A)=\tr(\delta \rho)=0$, we have
\begin{equation}
    S[\rho+\delta \rho]=S[\rho_A]+\frac{\beta}{N_A} \tr_A(H_A \delta\rho_A)\,.
\end{equation}
Hence the variation of entropy density of sub-system $A$ is
\begin{equation}
    \delta S=i\frac{\beta}{N_A}\int_0^t \tr_A( H_A \tr_{\bar{A}}(e^{is H}[\delta H,\rho]e^{-isH}))\D{s}\,.
\end{equation}
If $O_A$ is an operator with support in $A$ and $O$ an arbitrary operator, we have $\tr_A(O_A \tr_{\bar{A}}(O))=\tr(O_A O)$. Hence we can write, reorganizing the commutator and using again that $\rho$ commutes with $H$
\begin{equation}
    \delta S=i\frac{\beta}{N_A}\int_0^t \tr( \delta H e^{-isH}[\rho,H_A]e^{isH})\D{s}\,.
\end{equation}
The commutator $[\rho,H_A]$ is of order $|\partial A|$, since $\rho$ approximately commutes with $H$ in the bulk of $A$. Hence we have the change of entropy density
\begin{equation}\label{ssubleading}
    \delta S= \mathcal{O}\left(t \frac{|\partial A|}{N_A}\delta H\right)\,.
\end{equation}
It follows that in the entropy density, the term of order $t\delta H$ vanishes in the thermodynamic limit, when $|\partial A|/N_A\to 0$. All the higher order terms $t\delta H^n$ with $n>1$ vanishes in the adiabatic limit, that is rescaling $t\to \lambda t$ and $\delta H\to \delta H/\lambda$, with $\lambda\to\infty$. In \eqref{correction} and subsequently in the expansion of $S[\rho+\delta \rho]$, the second-order term in the $\delta H$ expansion is a priori of order $\mathcal{O}(t^2)$ with time. A term $t^2 \delta H^2$ would give a finite contribution in the adiabatic limit $\lambda\to\infty$. However, we show in the Appendix \ref{appendix} that after closer inspection, the order $\delta H^2$ scales as $\mathcal{O}(t)$ with time, the leading order $\mathcal{O}(t^2)$ disappearing when the unperturbed density matrix $\rho$ is a Gibbs state. Hence these higher-order terms do vanish in the adiabatic limit. Hence, the evolution is isentropic in the adiabatic limit.

To get the particular result \eqref{finalbeta} in the case of $H_0$ given by 
\eqref{h0}, we must compute the initial entropy density as a function of $\beta_0$. The Gibbs state of $H_0$ in \eqref{h0} at inverse temperature $\beta_0$ is
    \begin{equation}
    \begin{aligned}
        \rho_0&=\bigotimes_{j=1}^N \frac{e^{\beta_0}\frac{I-X_j}{2}+e^{-\beta_0}\frac{I+X_j}{2}}{e^{\beta_0}+e^{-\beta_0}}\\
        &=\bigotimes_{j=1}^N \left(\frac{I}{2}+\frac{\tanh\beta_0}{2}X_j \right)\,.
    \end{aligned}
    \end{equation}
    From this we compute the entropy density of the initial state
    \begin{equation}
    \begin{aligned}
        S_0(\beta_0)&=-\frac{1+\tanh\beta_0}{2}\log \frac{1+\tanh\beta_0}{2}\\
        &-\frac{1-\tanh\beta_0}{2}\log \frac{1-\tanh\beta_0}{2}\,.
    \end{aligned}
    \end{equation}
    We note that in this particular case, computing the entropy of the entire system $\mathcal{S}[\rho]$ or only the entropy density on half of the system as in \eqref{sdensity} gives the same value. By computing $\partial_{\beta_0}S_0(\beta_0)$ we obtain \eqref{finalbeta}.

    Let us make a few side remarks on the isentropicity of the adiabatic evolution. Firstly, this constant entropy density only holds when the entropy is computed on a reduced density matrix on a large number of sites (and such that the boundary of this subset is much smaller than the bulk). In particular, the entropy of a one-site reduced density matrix is not constant, even in the adiabatic limit. Secondly, the well-known area-law of entanglement entropy for ground states of local Hamiltonians \cite{eisert2008area} has a simple interpretation in terms of this isentropicity. If we start from the ground state of the initial Hamiltonian, the initial entropy density is $0$ (because it is a product state), so according to the isentropicity of adiabatic evolution in the thermodynamic limit, the entropy \emph{density} of the final state obtained after adiabatic evolution, namely of the ground state of the final Hamiltonian, must vanish in the thermodynamic limit. This implies that the entanglement entropy of this ground state must scale slower than $N_A$ as $N_A\to\infty$, namely the ground state cannot have volume-law entanglement. The sub-leading scaling $|\partial A|/N_A$ in \eqref{ssubleading} is compatible with an area-law entanglement entropy for this ground state, namely an entanglement entropy that scales as $|\partial A|$.

\subsection{Numerical simulations and checks}

\begin{figure}
    \centering
    \includegraphics[width=\linewidth]{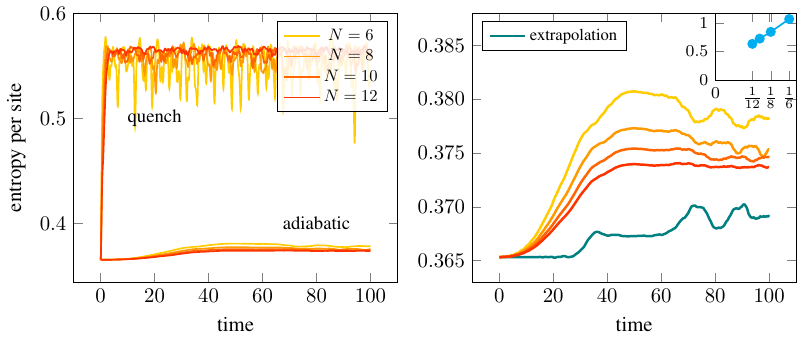}
    \caption{Entropy per site as a function of time, for $\beta=1$, using Trotter steps ${\rm d}t=0.05$, in a quench setting and in an adiabatic setting, for different system sizes $N$. The entropy density \eqref{sdensity} is calculated on the first $N/2$ sites. The right panel shows a zoomed-in version of the curves in the adiabatic setting, as well as a linear extrapolation in $1/N$. The inset shows the area between the curves and the constant line corresponding to the initial entropy density, as a function of $1/N$.}
    \label{fig:adiabatic}
\end{figure}

In Fig \ref{fig:adiabatic}, we show numerical simulations comparing the time evolution of the entropy in a quench setting and in an adiabatic setting. We start at time $t=0$ in a Gibbs state of $H_0$ in \eqref{h0} with $\beta=1$. In the quench setting, we time-evolve this initial density matrix with the 1D Ising Hamiltonian 
\begin{equation}\label{1dising}
    H_f=J\sum_{j=1}^N Z_j Z_{j+1}+h_x \sum_{j=1}^N X_j+h_z \sum_{j=1}^N Z_j\,,
\end{equation}
for $J=-1,h_x=h_z=1$ and periodic boundary conditions. We compute the entropy density of the reduced density matrix on the first $N/2$ sites, as defined in \eqref{sdensity}. We see that the entropy per site quickly grows to saturate up to oscillations at some fixed value independent of system size, in agreement with standard results of entanglement entropy growth after a quantum quench \cite{calabrese2005evolution}. In the adiabatic setting, we time-evolve the initial density matrix with the Hamiltonian $H(t)$ according to \eqref{pathadiabatic}. In that case, we see that the growth of the entanglement entropy is much milder, and decreases with system size. Since the adiabatic evolution should be isentropic only in the thermodynamic limit $N\to\infty$, the slight growth in finite size is not incompatible with our result, and is contained in the sub-leading terms in \eqref{ssubleading}. In order to estimate the entropy density growth in the thermodynamic limit, we perform a linear extrapolation with $1/N$. We see that the value obtained at $N=\infty$ is well compatible with an entropy per site that remains constant throughout the evolution. 

\begin{figure}
    \centering
    \includegraphics[width=0.49\linewidth]{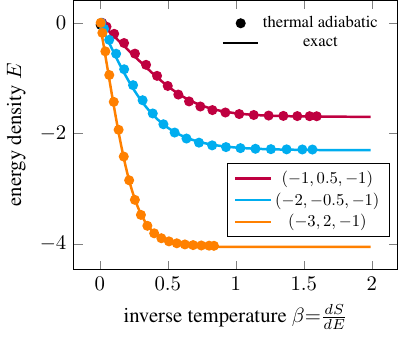}\includegraphics[width=0.51\linewidth]{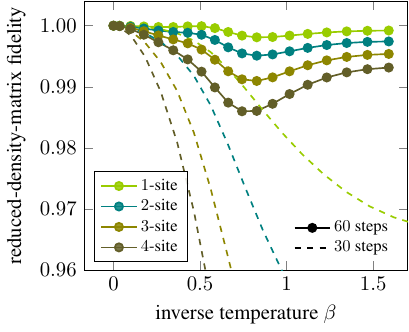}
    \caption{\textit{Left panel:} Energy density as a function of final inverse temperature $\beta_f$ \eqref{finalbeta}, using the thermal state preparation (bullets) and using the exact Gibbs state $e^{-\beta H_f}$ (lines), in 1D Ising models \eqref{1dising} in size $N=12$ for parameters $(h_x,h_z,J)$ indicated in the legend. The adiabatic state preparation uses $60$ Trotter steps of size $0.1$. \textit{Right panel:} fidelity between reduced density matrices on $1,2,3,4$ adjacent sites, computed with the thermal adiabatic approach with $30$ or $60$ Trotter steps, and computed from the exact Gibbs state with same temperature, with the same case $(h_z,h_x,J)=(-1,0.5,-1)$ in size $N=12$ as left panel.}
    \label{fig:comparison}
\end{figure}

As further check of our result, we now numerically test the adiabatic preparation of local thermal states with this procedure. We start from a Gibbs state of $H_0$ in \eqref{h0} at inverse temperature $\beta_0$, and perform an adiabatic evolution between $H_0$ and a 1D Ising model given by \eqref{1dising}. We perform $60$ Trotter steps of size $0.1$, measure the energy $E$, and compute the final inverse temperature $\beta$ with formula \eqref{finalbeta}. This simulation is done with exact density matrix evolution. We plot in the left panel of Fig \ref{fig:comparison} the resulting curve $E(\beta)$ for different values of $h_x,h_z,J$, and compare with the energy computed in the Gibbs state $\propto e^{-\beta H_f}$. We observe very good agreement between the two approaches. Since our approach prepares states that are only locally thermal, the density matrix on the entire system does not have to be close to a Gibbs state on the entire system. However, reduced density matrices should be close. In the right panel of Fig \ref{fig:comparison}, we plot in the same setting as before the fidelity between the reduced density matrices obtained from the thermal adiabatic evolution and from the exact Gibbs state, considering $1,2,3$ or $4$ adjacent sites, computed for two different number of Trotter steps $30$ and $60$. We see that the fidelity increases when doing a slower adiabatic evolution, as expected. The fidelities obtained for the slower evolution are very close to $1$, confirming our claims.

\section{Effect of noise on thermal state preparation} 

\subsection{Measuring entropy on hardware}

We now would like to take into account the effect of noise on the thermal adiabatic state preparation. Namely, we will consider the effect of errors applied after every gate, with a certain error rate. In the adiabatic limit, there are more and more gates to implement in the circuit, so that to get a non-trivial adiabatic limit in the presence of noise, one has to scale the error rate $p$ per gate as $p\propto 1/T$. In that case, the amount of noise injected in the system in a time $\delta t/T$ is small. From the thermalization property, one thus has that the system is at thermal equilibrium at all times. The ``noisy'' inverse temperature $\tilde{\beta}_f$ can thus be computed as
\begin{equation}\label{beta1tilde}
    \tilde{\beta}_f=\frac{\partial_{\beta_0}\tilde{S}_f}{\partial_{\beta_0}\tilde{E}_1}\,,
\end{equation}
with $\tilde{S}_f,\tilde{E}_f$ the entropy and energy densities in the final state. However, contrary to the noiseless case, entropy density is \emph{not} conserved in the presence of noise. The entropy in the end state cannot thus be computed from the entropy in the initial state, namely in the presence of gate noise we have $\partial_{\beta_0}\tilde{S}_f \neq \partial_{\beta_0}S_0$.

To evaluate the increase in entropy during the adiabatic evolution, let us modify the circuit as follows. Instead of doing the entire adiabatic evolution from $t=0$ to $t=T$, we do the evolution only up to $t=T/2$, and then a \emph{backward} time evolution down to $t=0$, exactly inverting the gates applied in the first half of the circuit. In the noiseless setting, the state of the system at the end of the circuit is thus exactly the same as at the beginning. In the noisy setting, the system does not exactly come back to the initial state. This is a very standard ``mirror circuit'' used for benchmarking quantum computer noise \cite{proctor2022measuring}. However, in this adiabatic context, additional interpretation can be made of this benchmark protocol. In the adiabatic limit $T\to\infty$, since the path implements an adiabatic evolution back to the initial Hamiltonian, we arrive at a thermal density matrix of the initial Hamiltonian. Because the thermal states of the initial Hamiltonian \eqref{h0} are product states with an on-site density matrix equal to
\begin{equation}
    \frac{I}{2}+\frac{\tanh \beta}{2}X\,,
\end{equation}
with some inverse temperature $\beta$, the temperature of the thermal state in which the state is can be measured by simply measuring the $X$ observable on the sites. From the temperature, one can deduce the value of the entropy density in the state. So, assuming that the state is in a thermal state of $H_0$ in \eqref{h0}, we have the entropy density
\begin{equation}\label{eqm}
    S=-\frac{1+\langle X\rangle}{2}\log \frac{1+\langle X\rangle}{2}-\frac{1-\langle X\rangle}{2}\log \frac{1-\langle X\rangle}{2}\,.
\end{equation}
In the absence of noise, by computing the entropy density this way, we will find  exactly the initial value of entropy density. In the presence of noise, the entropy will have increased at the end of the circuit. Let us assume that in the presence of noise, the measured $X$ observable is
\begin{equation}\label{eqmbeta0}
    m=r \tanh \beta_0\,,
\end{equation}
with $r>0$ a noise amplitude, equal to $r=1$ in the absence of noise, and $r<1$ in the presence of noise. The entropy density with noise is thus
\begin{equation}\label{eqmnoisy}
\begin{aligned}
    \tilde{S}=&-\frac{1+m}{2}\log \frac{1+m}{2}-\frac{1-m}{2}\log \frac{1-m}{2}\,.
\end{aligned}
\end{equation}
We now make the assumption that this is the entropy density in the final state $\tilde{S}_f$ when the original adiabatic evolution is performed. Namely, we make the assumption that the entropy created by the imperfect gates is independent of whether we implemented a forward or backward evolution in the second half of the circuit. Since these two settings contain almost identical gates that differ only by the gate angles, this is a reasonable assumption. From this assumption, we can compute the inverse temperature $\tilde{\beta}_f$ that we obtain at the end from \eqref{beta1tilde}. We find
\begin{equation}
    \partial_{\beta_0}\tilde{S}=\frac{r(1-r^2\tanh^2\beta_0) \argtanh(r\tanh\beta_0)}{1+(1-r^2)\sinh^2\beta_0}\,.
\end{equation}
The precise value of the final energy density $E_f$ will depend on the model. However, at large $\beta_0$, the initial state can be interpreted as being $|+...+\rangle$ with probability $1-e^{-2\beta_0}+\mathcal{O}(e^{-4\beta_0})$, and some other states with probability $e^{-2\beta_0}+\mathcal{O}(e^{-4\beta_0})$. Similarly, at low noise $1-r\to 0$, the final state is the noiseless state with probability $\mathcal{O}(r)$ and some other state with probability $\mathcal{O}(1-r)$. Hence we can write
\begin{equation}
    \partial_{\beta_0} \tilde{E}_1=C e^{-2\beta_0}+\mathcal{O}(e^{-4\beta_0})+\mathcal{O}(1-r)\,,
\end{equation}
with some model-dependent constant $C>0$. At small noise rate $1-r\to 0$ and large $\beta_0$, we thus get
\begin{equation}
    \tilde{\beta}_f=\frac{4\argtanh(r\tanh\beta_0)}{C}\,.
\end{equation}
In particular, in the presence of noise we have $r<1$, and so there is a maximal inverse temperature $\tilde{\beta}_f$ that can be attained, that scales as
\begin{equation}\label{scaletildebeta}
    \tilde{\beta}_f^{\max} \sim \frac{2\log (1-r)}{C}\,.
\end{equation}
This is an important difference with the noiseless, ideal case: While a perfect hardware would be able to prepare local thermal states for the final Hamiltonian with any inverse temperature $\beta_f$, only inverse temperatures $\beta_f<\tilde{\beta_f}^{\rm max}$ can be prepared on a noisy hardware. This logarithmic dependence of the best possible temperature with the error rate of the quantum computer, notoriously difficult to lower down, might look pessimistic. However, at  temperature lower than the gap of the system, all physical quantities will have exponentially small deviations from the ground state value. Since noise can always be seen as a certain probability to be in an excited state, the incurred increase of energy naturally translates into an inverse temperature that is only logarithmically large with the error rate. 

Let us come back to the assumption that the entropy of the system can be obtained with \eqref{eqm} by measuring $m$, the expectation value of $X$ after the mirror adiabatic evolution. Generally, we expect expectation values of local observables to decrease exponentially with the depth of a noisy circuit, when noise is for example depolarizing. Hence, if this assumption is correct, the entropy density as a function of time $t$ along an adiabatic evolution should follow the curve
\begin{equation}\label{adiabaticcurve}
    \begin{aligned}
        S(t)=&-\frac{1+e^{-\alpha t}\tanh\beta_0}{2}\log \frac{1+e^{-\alpha t}\tanh\beta_0}{2}\\
        &-\frac{1-e^{-\alpha t}\tanh\beta_0}{2}\log \frac{1-e^{-\alpha t}\tanh\beta_0}{2}\,,
    \end{aligned}
\end{equation}
with $\alpha$ some fixed noise-dependent parameter.

\begin{figure}
    \centering
    \includegraphics[width=\linewidth]{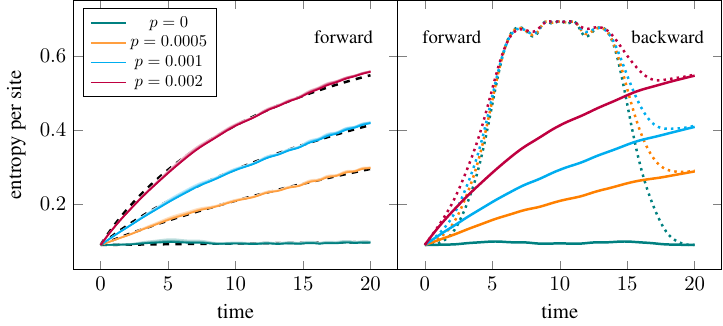}
    \caption{\emph{Left panel:} entropy per site  \eqref{sdensity} as a function of time in an adiabatic setting, starting from $\beta=2$, with Trotter steps ${\rm d}t=0.1$, for different system sizes (from light $N=6$ to dark $N=12$, essentially superimposed), for different single-qubit depolarizing channel amplitude $p$ per Trotter step. The dashed black lines show the value of the model \eqref{adiabaticcurve} after fitting $\alpha$. \emph{Right panel:} continuous lines are obtained in the same setting as left panel for $N=12$, but with a backward (mirror circuit of forward) adiabatic evolution starting at time $10$. The dotted lines indicate the entropy density computed with \eqref{eqm}.}
    \label{fig:adiabatic_noise}
\end{figure}

In Fig \ref{fig:adiabatic_noise} we numerically test these findings. We perform an adiabatic evolution from Hamiltonian $H_0$ in \eqref{h0} to Hamiltonian \eqref{1dising} with $J=-1$, $h_x=h_z=1$, starting from a Gibbs state at $\beta=2$, and modeling the noise by the application of a single-qubit depolarizing channel with amplitude $p$ on every qubit after every Trotter step. We again measure the entropy density of the system by measuring the entropy of the reduced density matrix on the first $N/2$ sites. We see in the left panel that the entropy density steadily increases with time in the presence of noise. Fitting the parameter $\alpha$ of \eqref{adiabaticcurve}, we see that \eqref{adiabaticcurve} accounts well for the growth of entropy density in the noisy setting. In the right panel of Fig \ref{fig:adiabatic_noise}, we then show the entropy density as a function of time in the mirror adiabatic circuit, and show in dotted lines the entropy density computed with formula \eqref{eqm} along the path. We see that the entropy density at the end of the mirror circuit is very similar to that of the (only-forward) adiabatic evolution. Moreover, formula \eqref{eqm} agrees perfectly at the beginning and the end of the adiabatic evolution, while disagreeing in the middle, which is expected since the system is not in a product state. This confirms that the entropy density in the presence of noise at the end of the adiabatic evolution can be evaluated by performing a mirror adiabatic evolution circuit, measure $X$, and use formula \eqref{eqm}

\subsection{Adiabatic thermal state preparation is noise-resilient} 

 We now state a protocol to prepare thermal states of a Hamiltonian $H$. 

\begin{enumerate}
    \item For a given $\beta_0$, we prepare the quantum computer in a Gibbs state of $H_0$ in \eqref{h0}. We evolve the state of the quantum computer with the time-dependent Hamiltonian $(1-\frac{t}{T})H_0+\frac{t}{T} H_f$, for a given adiabatic time $T$ assumed to be large enough so that the evolution is adiabatic. We then measure the energy of the final Hamiltonian $H_f$ obtained, denoted $E(\beta_0)$.
    \item For the same $\beta_0$, we prepare again the quantum computer in the Gibbs state of $H_0$. We implement the same adiabatic evolution, but only up to half of the time $t=T/2$, and then apply the exact inverse circuit, so as to come back exactly to the initial state for an ideal noiseless circuit. We measure $m$ the average expectation value of $X$ on the qubits, and deduce the entropy density $S(\beta_0)$ through formula \eqref{eqm}.
    \item Collecting the values of $E(\beta_0)$ and $S(\beta_0)$ for several values of $\beta_0$, we compute an estimated final temperature $\beta(\beta_0)=\frac{dS(\beta_0)}{dE(\beta_0)}$ when varying $\beta_0$. We deduce the curve $E(\beta)$.
\end{enumerate} 
If we assume that the measured $m=\langle X\rangle$ depends on $\beta_0$ only through \eqref{eqmbeta0} with a fixed $r$, then it may suffice to perform step $2$ of this protocol only for one value of $\beta_0$, and deducing the curve $S(\beta_0)$ from \eqref{eqmbeta0} and \eqref{eqm}. As stated before, the protocol also applies to other simple initial Hamiltonians $H_0$, provided the entropy measurement of formula \eqref{eqm} can be adapted accordingly, namely if the entropy of Gibbs states of $H_0$ can be measured by just measuring some simple local observable.

\begin{figure}
    \centering
    \includegraphics[width=\linewidth]{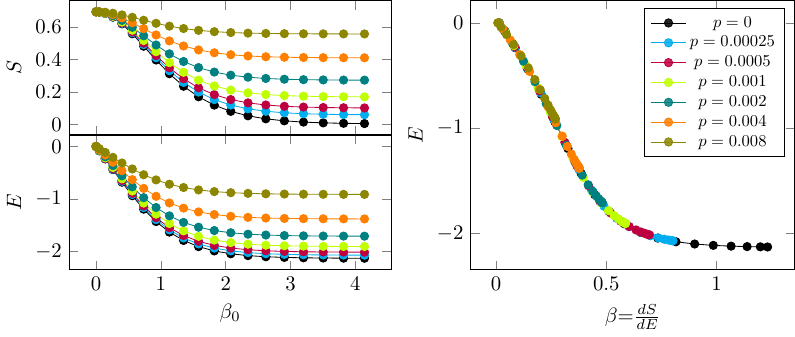}
    \caption{\emph{Left panel:} Energy density $E$ and entropy density $S$ \eqref{sdensity} as a function of initial inverse temperature $\beta_0$, using the thermal state preparation for 1D Ising with fields $h_x=-1$, $h_Z=1$, $J=-1$, in size $N=12$, with $60$ Trotter steps of size ${\rm d}t=0.1$, for different noise levels $p$. Noise is modeled by rounds of single-qubit depolarizing channels on every qubit after each Trotter step. The entropy density $S$ is estimated through the mirror protocol and Eq \eqref{eqm}, which is implementable on hardware. \emph{Right panel:} energy density as a function of $\beta=\frac{dS}{dE}$ computed from the values of the left panel as in \eqref{beta1tilde}.}
    \label{fig:noise}
\end{figure}

In the left panel of Fig \ref{fig:noise}, we plot the curves $S(\beta_0)$ and $E(\beta_0)$ obtained with this protocol in the 1D Ising model \eqref{1dising} in size $N=12$, for different noise levels in the time evolution. We chose a relatively small value of $N$ so as to be able to completely remove statistical noise in the noisy simulations by performing density matrix evolution. We observe that, as expected, the value of energy reached at large $\beta_0$ (i.e. when starting from the ground state of $H_0$ and implementing the usual adiabatic evolution) increases significantly with the amount of noise. Similarly, the entropy density increases too with noise, from $S=0$ in absence of noise to $S=\log 2$ for maximal noise. However, when plotting $E$ as a function of $\beta=\frac{dS}{dE}$ in the right panel, we observe that remarkably, the different curves at different noise levels \emph{exactly collapse} onto a same curve. We only observe that the maximal $\beta$ reachable with this procedure decreases when increasing noise, in a way that is in agreement with the scaling \eqref{scaletildebeta} for the maximal inverse temperature. This shows that the thermal adiabatic state preparation is noise-resilient, in the sense that a point on the $E(\beta)$ plot obtained is insensitive to noise rate. 

This numerical result holds for the special case of depolarizing noise, which is a simplification of the real noise occurring on actual hardware. However, several other types of noise, such as biased Pauli noise, coherent noise or leakage errors, can be transformed into depolarizing noise with e.g. Pauli twirling and systematic leakage repump \cite{wallman2016noise,hayes2020eliminating}. Observing a perfect collapse for depolarizing noise is thus a meaningful case.

\section{Effect of lack of adiabaticity on thermal state preparation} 
 Up to now, we have considered a slow enough time evolution, i.e. a large enough adiabatic time $T$, so that the time evolution can be considered adiabatic with a good approximation. This adiabaticity is important for arguing that (i) the state prepared is a thermal state, and (ii) the increase of entropy due to noise can be measured with the mirror circuit. In practice however, when implementing the protocol on an actual quantum computer, one is constrained in the number of gates that can be implemented, and the resulting evolution may significantly deviate from an adiabatic evolution. The protocol above to compute the $E(\beta)$ curve can still be implemented, and will give qualitatively similar curves, but quantitatively incorrect. Hence, it is valuable to have a protocol to test the adiabaticity of an evolution on the quantum computer.

\begin{figure}
    \centering
    \includegraphics[width=0.49\linewidth]{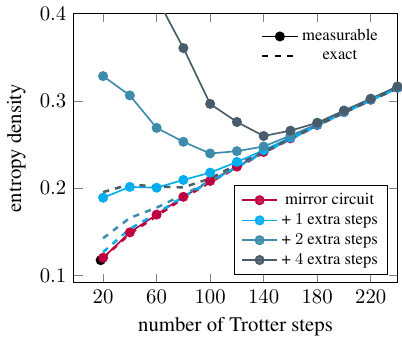}
    \includegraphics[width=0.49\linewidth]{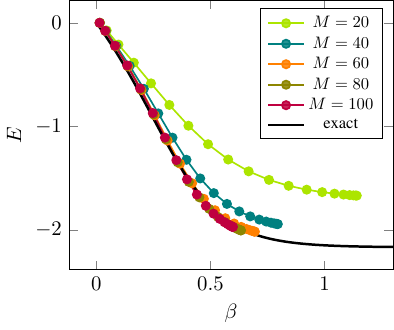}
    \caption{Left panel: entropy density measured through the expectation value of $X$ with formula \eqref{scaletildebeta}, at the end of a mirror circuit $U_M U_M^\dagger$ (purple) and when inserting $1,2$ or $4$ extra steps at the middle $U_M V U_M^\dagger$ (cyan), as a function of the total number of Trotter steps implemented in the circuit divided by $2$, for a noise rate $0.0005$ per Trotter step. The exact value of the entropy density is indicated by the dashed lines. Right panel: $E(\beta)$ curve obtained with the same conditions as Fig \ref{fig:noise}, for noise rate $p=0.0005$, for different number of Trotter steps $M$ in the adiabatic evolution.}
    \label{fig:adiabaticity}
\end{figure}

Let us consider $U_M$ the unitary operator implementing the adiabatic evolution from $H_0$ to $H_f$, using $M$ Trotter steps of a fixed size ${\rm d}t$. When starting from a Gibbs state of $H_0$, and implementing the exact mirror circuit $U_M U_M^\dagger$, one obtains back the same Gibbs state of $H_0$. For $H_0$ given by \eqref{h0}, the entropy density of this thermal state can be computed exactly by measuring $X$ on the qubits. In the presence of noise, the implementation of the mirror circuit $U_M U_M^\dagger$ will result in an operation that differs from identity. We saw previously in Fig \ref{fig:adiabatic_noise} that measuring $X$ on the qubits after this noisy mirror circuit still gives an excellent estimate of the entropy density of the resulting noisy state. In the left panel of Fig \ref{fig:adiabaticity}, we test how robust this fact is to the absence of adiabaticity. We implement this mirror circuit in the presence of noise, for different number of Trotter steps, in such a way that the evolution is far from adiabatic at low Trotter step number $M$, and a good approximation of the adiabatic evolution for the largest values of $M$. We compare then the entropy density estimated by measuring $X$, to the exact entropy density in the resulting state. We see that even for very fast evolutions where adiabaticity is not satisfied, measuring $X$ still gives a very good estimate of the entropy density in the state. 

Now, we make the following simple remark. If the evolution is adiabatic, the resulting state after application of $U_M$ should be a thermal state of the final $H_f$. Evolving this state with $H_f$ should thus leave the state invariant. We modify thus the circuit by inserting between $U_M$ and the inverse $U_M^\dagger$ a few Trotter steps of the final Hamiltonian $H_f$, whose unitary we denote $V$. Although differing from the identity operator, the circuit $U_M V U_M^\dagger$ should, if the evolution is adiabatic, map the initial thermal state of $H_0$ to another thermal state of $H_0$. If the number of Trotter steps done in $V$ is small, gate noise incurred by these extra steps will only slightly increase the entropy density. Hence, if the evolution is adiabatic, (i) the entropy density of the resulting state should be similar whether we insert between $U_M$ and $U_M^\dagger$ the extra Trotter steps $V$ or not, and (ii) in both cases the states are thermal and one should be able to measure the entropy density by measuring $X$. On the contrary, if the evolution is \emph{not} adiabatic, then the state after application of $U_M$ differs from a thermal state of $H_f$. It will in general be modified by the application of Trotter steps of $H_f$. This will, generically for a quantum quench, increase the entanglement entropy of the state. Moreover, after being mapped back with $U_M^\dagger$, the resulting state has no reason to be thermal. We expect thus the measurement of $X$ to be different from the actual entropy of the state.

In the left panel of Fig \ref{fig:adiabaticity}, we show both the exact entropy density after the circuit $U_M V U_M^\dagger$, and the entropy density estimated by measuring $X$. We observe that (i) the two differ significantly at small number of Trotter steps, and (ii) the exact entropy density differ in the mirror circuit $U_M U_M^\dagger$ and in $U_M V U_M^\dagger$ at small number of steps, but agree at large number of steps. On a quantum computer, only $X$ can be measured, not the exact entropy density. To test how much the agreement of the measurement of $X$ in $U_M U_M^\dagger$ and in $U_M V U_M^\dagger$ correlates with adiabaticity, we show in the right panel of Fig \ref{fig:adiabaticity} the curves $E(\beta)$ obtained for different number of steps. If the evolution is adiabatic for a given number of steps $M$, then increasing $M$ should not significantly change the curve. We observe indeed that the curves for $M=80,100$ start to essentially superimpose (up to the smaller maximal $\beta$ for larger values of $M$, because of noise). This corresponds indeed to values in the left panel of Fig \ref{fig:adiabaticity} where with one extra step, the two measured entropy densities start to coincide. In contrast, for $M=20,40,60$, the curves in the right panel of \ref{fig:adiabaticity} significantly differ, and so do the measured entropy densities in the left panel.

\section{Quantum computing hardware implementation}
\subsection{Setup}
 We now present an implementation of our protocol on actual quantum computing hardware. We consider a 2D square lattice of dimensions $L_x=5$, $L_y=4$, and define the Ising model
\begin{equation}
    H=-\sum_{\langle i,j\rangle}Z_i Z_j+h_x \sum_{j=1}X_j\,,
\end{equation}
where $\langle i,j\rangle$ means that sites $i,j$ are neighbours on the lattice, with periodic boundary conditions. We set $h_x=2$. The adiabatic path is taken to be \eqref{pathadiabatic} with the initial Hamiltonian $H_0$ therein being $-H_0$ in \eqref{h0}. We trotterize this adiabatic evolution into a unitary operator $U$ that contains $M$ Trotter steps
\begin{equation}
    U=V(J_{M-1},f_{M-1})V(J_{M-2},f_{M-2})...V(J_0,f_0)\,,
\end{equation}
with the Trotter step operator
\begin{equation}
    V(J,f)=e^{if\sum_j X_j}e^{iJ\sum_{\langle j,k\rangle} Z_jZ_k}\,.
\end{equation}
We set the sequence
\begin{equation}
    f_n=\frac{1}{(a+b n)^c}\,, \quad J_n=-\frac{f_n}{h_x}\left(\frac{n+\frac{1}{2}}{M}\right)^d\,,
\end{equation}
with a number of steps $M=16$, and optimize the parameters $a,b,c,d$ to
\begin{equation}
\begin{aligned}
    &a=1.186\,,\qquad b=0.077\,,\\
    &c=2.181\,,\qquad d=0.469\,.
\end{aligned}
\end{equation}
 The energy density of $H$ obtained with a perfect implementation of $U$ is $-2.465$. As a comparison, the energy density of the three lowest energy states are $-2.51147$, $-2.51125$ and $-2.26652$. The mirror circuit used in the protocol is similarly defined as
 \begin{equation}
 \begin{aligned}
    U_{\rm mirror}=&V^\dagger(J_{0},f_{0})V^\dagger(J_{1},f_{1})...\\
    &...V^\dagger(J_{M/2-1},f_{M/2-1})V(J_{M/2-1},f_{M/2-1})...\\
    &...V(J_0,f_0)\,,
 \end{aligned}
\end{equation}
with the same parameters as for $U$. On hardware, the operators $U$ and $U_{\rm mirror}$ are each implemented with respectively $640$ two-qubit gates. 

At large initial inverse temperature $\beta_0$, the thermal states of $H_0$ are
 \begin{equation}
 \begin{aligned}
     \rho=&\left(\frac{1+\tanh \beta_0}{2}\right)^{N}|-...-\rangle\langle -...-|\\
     &+\frac{1-\tanh \beta_0}{2}\left(\frac{1+\tanh \beta_0}{2}\right)^{N-1}\\
     &\qquad\qquad\times \sum_{j=1}^N |-...\underset{j}{+}...-\rangle \langle -...\underset{j}{+}...-|\\
     &+\mathcal{O}\left(\left(\frac{1-\tanh \beta_0}{2}\right)^2\right)\,,
 \end{aligned}
 \end{equation}
where in the third line there is a single $+$ at site $j$. Since $H$ and $U$ are translation invariant, the energy and the average of any local observable over all the sites is the same in all the states $U|-...\underset{j}{+}...-\rangle$. Let us thus denote $e$ the energy density of $U|-...-\rangle$, $e'$ the energy density of $U|+-...-\rangle$, $m$ the average value of $X$ of the mirror circuit $U_{\rm mirror}|-...-\rangle$, and $m'$ the average value of $X$ of the mirror circuit $U_{\rm mirror}|+-...-\rangle$, all these quantities being measured on hardware. Then the inverse temperature $\beta$ with respect to $H$ of the state $U|-...-\rangle$ prepared on hardware is
\begin{equation}\label{formula}
    \beta=\frac{1}{2} \frac{m-m'}{e-e'}\log \frac{1-m}{1+m}\,.
\end{equation}

\subsection{Hardware results}
We implemented this protocol on Quantinuum's H1-1 quantum computing device. This is an ion-trap device hosting $20$ qubits, with a native entangling gate $e^{i\theta ZZ}$ operating with fidelity around $99.9\%$. We compute $m$ and $m'$ with $1000$ shots each, and $e$ and $e'$ with $3000$ shots each, distributed as $1500$ shots measured in the $Z$ basis and $1500$ shots measured in the $X$ basis. 

The hardware results are presented in Table \ref{results_hardware}. The energy density of the state prepared is measured to be
\begin{equation}
    e=-2.334\pm 0.0186\,.
\end{equation}
We measure an entropy per site
\begin{equation}
    S=0.1665\pm 0.0045\,,
\end{equation}
at the end of the circuit with $640$ two-qubit gates. Using \eqref{formula}, we find a final temperature $T=1/\beta$ given by
\begin{equation}
    T=2.562  \pm 0.256 \,.
\end{equation}
The error bar on the inverse temperature $\beta$ displays significant asymmetry, which is why we present the results in terms of the temperature, where the error bar is symmetric. We recall that the gap between the quasi-twice-degenerate ground state and the first excited state is $4.895$. This temperature is thus relatively low, at around half of the gap.

\begin{table}
    \centering
    \begin{tabular}{|c|c||c|c|}
    \hline
    Quantity & Hardware & Noiseless& \begin{tabular}{@{}c@{}}Exact \\ $\beta_0=1.6$\end{tabular}  \\
    \hline
    \hline
      $\langle ZZ\rangle$&  $1.184   \pm 0.0143$  & $1.417$ & $1.06$\\
     $\langle X\rangle$&  $-0.575   \pm 0.00597$  & $-0.524$& $-0.58$\\
     $e$&  $-2.334   \pm 0.0186$  & $-2.465$& $-2.21$\\
          \hline
    \hline
    $\langle ZZ\rangle'$&  $0.745  \pm 0.0123$  & $0.886$& $0.73$\\
    $\langle X\rangle'$&  $-0.641  \pm 0.00518$  & $-0.623$& $-0.60$\\
      $e'$&  $-2.0279   \pm 0.0161 $ & $-2.132$& $-1.94$\\
       \hline
    \hline
      $m$&  $-0.921 \pm 0.0028$ & $-1$& $-0.92$\\
       \hline
       \hline
      $m'$&  $-0.846 \pm 0.0026$ & $-0.9$& $-0.83$\\
       \hline
       \hline
       entropy density&  $0.1665 \pm 0.0045 $ & $0$& $1.68$\\
    temperature&  $2.562  \pm 0.256  $ & $0$& $1.91$\\
       temperature/gap&  $0.523  \pm 0.0524  $ & $0$& $0.39$\\
       \hline
    \end{tabular}
    \caption{Results of the thermal adiabatic evolution. The ``hardware'' column shows the results obtained from Quantinuum's H1-1 ion-trap machine. The ``noiseless'' column shows the exact results that we would obtain by running the same circuits without noise. The ``exact $\beta_0=1.6$'' column shows the values obtained for noiseless simulations, but when starting from a thermal state of $-H_0$ with inverse temperature $\beta_0=1.6$.  $\langle ZZ\rangle$ denotes the expectation value of $\frac{1}{N}\sum_{\langle i,j\rangle}Z_iZ_j$, and $\langle X\rangle$ that of $\frac{1}{N}\sum_{j}X_j$. The values without prime correspond to starting from $|-...-\rangle$, and the values with prime to starting from $|+-...-\rangle$. We show the deduced values for entropy density and temperature measured on hardware.}
    \label{results_hardware}
\end{table}

Let us analyze the hardware results in the light of the discussion of our protocol. As described above, with the mirror circuit approach, we measure on hardware an entropy per site $0.1665$ when starting from an infinite inverse temperature $\beta_0=\infty$, namely when starting from the ground state of $-H_0$. According to our arguments, if the time evolution is adiabatic and faithfully implemented, the energy measured on hardware should correspond to the energy of a noiseless evolution, but starting from an initial thermal state of $-H_0$ with entropy per site $0.166$. This corresponds to an initial inverse temperature $\beta_0=1.6$. When performing such noiseless time evolution, we find an energy $-2.22$. This significantly differs from the energy obtained on hardware, $-2.334\pm 0.0186$, by more than $6$ standard deviations. However, starting from such initial inverse temperature $\beta_0=1.6$, the application of our protocol gives a temperature $1.93$ as shown in Table \ref{results_hardware}, which differs from hardware by only $2.5$ standard deviations. While it is statistically likely that there is a bias, the pair of values of entropy and temperature are much more in agreement with hardware than the pair of values of entropy and energy.

\subsection{Interpretation of the results}

To understand the source of discrepancy, we can compare these results with noisy simulations. To measure the same value of $X$ as on hardware for the mirror circuit, we need to put a depolarizing channel with amplitude $1.1 \cdot 10^{-3}$ after every two-qubit gate. This is in very good agreement with component benchmarks of the H1-1 machine. The energy obtained with noisy simulations with this noise model is then $-2.24$. This energy matches well the noiseless simulations starting from $\beta_0=1.6$. Since adiabaticity and thermalization are required in our arguments to obtain these two values matching in the noisy numerics, we can reasonably conclude that these two properties are indeed approximately satisfied. The most likely explanation of discrepancy with hardware is thus a source of noise that \emph{does not} necessarily increase entropy. One such source of error is leakage error, where an ion hosting a qubit is excited into a state outside of the computational space. Such a leaked qubit is always measured as $1$. In terms of measurement outcomes, these leakage errors thus do not necessarily increase the entropy of the system: for example, in the limit where all the qubits leak, they are all measured in the $1$ state and there is zero entropy in the measurement outcomes. The fact that on the hardware results, the expectation value of $X$ after the application of $U$ is closer to $-1$ than in the noiseless case, both when starting from $|-...-\rangle$ and from $|+-...-\rangle$, suggests that leakage biases indeed the hardware results. 

Another source of errors that do not increase entropy is coherent noise. This coherent noise can in principle be systematically converted into incoherent noise through randomized compiling. This incoherent noise would then increase entropy and our protocol would apply. On Quantinuum's machines, memory error is a known source of coherent errors under the form of spurious magnetic fields in the $Z$ direction. However, taking the component benchmark value for memory error and adding spurious $Z$ rotations into simulations of the circuit, we did not observe significant changes in the results and were not able to explain the hardware results from this error source alone. Leakage error remains thus the most likely source of disagreement between energy and entropy measured on hardware. This source of error can be alleviated by using leakage detection gadget and post-selecting onto shots without any leakage events to test our understanding of the hardware results. However, these leakage detection gadgets require ancillas on the H1-1 machine, whereas all the $20$ qubits are used to describe the $5\times 4$ system studied. Future or bigger machines will be more adapted to study the effect of leakage and confirm our interpretation of the hardware results.

\section{Summary and conclusion}

 In this paper, we have discussed a way of preparing states on a quantum computer that are \emph{locally} thermal states of a given Hamiltonian. The method consists in initializing the state in a thermal state of a simple Hamiltonian, that can be exactly prepared efficiently, and to then time-evolve the state according to an adiabatic evolution identical to adiabatic ground state preparation. We showed that if thermalization holds, then the state obtained is locally at thermal equilibrium in the adiabatic regime. From the fact that the entropy density of local density matrices on a large sub-system is conserved during the adiabatic evolution in the thermodynamic limit, the energy/entropy curve can be drawn, and from it the energy/temperature curve. We studied the effect of depolarizing noise on the protocol, and showed that although the energy of a state is affected by noise on the hardware, the entropy is affected in a similar way, so that the energy/temperature curves obtained are robust to noise. These facts are the main results of the paper. Next, we proposed a protocol for testing the adiabaticity of an evolution. Finally, we implemented our protocol on Quantinuum's H1-1 ion-trap device, measuring the entropy and the temperature of a thermal state of the two-dimensional Ising model, yielding a benchmarking of the hardware in a concrete use case.

The work can be continued in multiple directions. The most interesting direction is to study what properties of the Hamiltonian can make the thermal state preparation slow down, similarly to a gap narrowing for ground state preparation. This is a completely open question. In particular, more precise runtime scalings of the algorithm with system size, temperatures or properties of the Hamiltonian should be investigated. Additionally, the fact that the adiabatic evolution on an actual quantum computer increases the entropy of local density matrices only through hardware noise, makes it a particularly interesting candidate protocol for noise benchmarking. Finally, it would be interesting to compare the efficiency of this thermal state preparation protocol with other protocols.

\subsection*{Acknowledgements}

 We thank Eric Brunner and Gabriel Matos for useful comments on the manuscript. The experimental data reported in this work were produced by the Quantinuum H1-1 quantum computer, Powered by Honeywell, in August 2025. 
 
\subsection*{Funding}
E.G. acknowledges support by the Bavarian Ministry of Economic Affairs, Regional Development and Energy (StMWi) under project Bench-QC (DIK0425/01).

\subsection*{Competing interests}
H.D. is a shareholder of Quantinuum.

\subsection*{Data availability}
The numerical and hardware data generated for this study are available upon request.

\subsection*{Author contributions}
All authors contributed to develop the method and main ideas. E.G. implemented the numerical analysis, hardware experiment and analytical results. E.G. wrote the first draft of the manuscript. All authors contributed to interpreting and discussing the results, and reviewed the manuscript.


\begin{thebibliography}{60}%
\makeatletter
\providecommand \@ifxundefined [1]{%
 \@ifx{#1\undefined}
}%
\providecommand \@ifnum [1]{%
 \ifnum #1\expandafter \@firstoftwo
 \else \expandafter \@secondoftwo
 \fi
}%
\providecommand \@ifx [1]{%
 \ifx #1\expandafter \@firstoftwo
 \else \expandafter \@secondoftwo
 \fi
}%
\providecommand \natexlab [1]{#1}%
\providecommand \enquote  [1]{``#1''}%
\providecommand \bibnamefont  [1]{#1}%
\providecommand \bibfnamefont [1]{#1}%
\providecommand \citenamefont [1]{#1}%
\providecommand \href@noop [0]{\@secondoftwo}%
\providecommand \href [0]{\begingroup \@sanitize@url \@href}%
\providecommand \@href[1]{\@@startlink{#1}\@@href}%
\providecommand \@@href[1]{\endgroup#1\@@endlink}%
\providecommand \@sanitize@url [0]{\catcode `\\12\catcode `\$12\catcode `\&12\catcode `\#12\catcode `\^12\catcode `\_12\catcode `\%12\relax}%
\providecommand \@@startlink[1]{}%
\providecommand \@@endlink[0]{}%
\providecommand \url  [0]{\begingroup\@sanitize@url \@url }%
\providecommand \@url [1]{\endgroup\@href {#1}{\urlprefix }}%
\providecommand \urlprefix  [0]{URL }%
\providecommand \Eprint [0]{\href }%
\providecommand \doibase [0]{http://dx.doi.org/}%
\providecommand \selectlanguage [0]{\@gobble}%
\providecommand \bibinfo  [0]{\@secondoftwo}%
\providecommand \bibfield  [0]{\@secondoftwo}%
\providecommand \translation [1]{[#1]}%
\providecommand \BibitemOpen [0]{}%
\providecommand \bibitemStop [0]{}%
\providecommand \bibitemNoStop [0]{.\EOS\space}%
\providecommand \EOS [0]{\spacefactor3000\relax}%
\providecommand \BibitemShut  [1]{\csname bibitem#1\endcsname}%
\let\auto@bib@innerbib\@empty
\bibitem [{\citenamefont {Dalzell}\ \emph {et~al.}(2023)\citenamefont {Dalzell}, \citenamefont {McArdle}, \citenamefont {Berta}, \citenamefont {Bienias}, \citenamefont {Chen}, \citenamefont {Gily{\'e}n}, \citenamefont {Hann}, \citenamefont {Kastoryano}, \citenamefont {Khabiboulline}, \citenamefont {Kubica} \emph {et~al.}}]{dalzell2023quantum}%
  \BibitemOpen
  \bibfield  {author} {\bibinfo {author} {\bibfnamefont {A.~M.}\ \bibnamefont {Dalzell}}, \bibinfo {author} {\bibfnamefont {S.}~\bibnamefont {McArdle}}, \bibinfo {author} {\bibfnamefont {M.}~\bibnamefont {Berta}}, \bibinfo {author} {\bibfnamefont {P.}~\bibnamefont {Bienias}}, \bibinfo {author} {\bibfnamefont {C.-F.}\ \bibnamefont {Chen}}, \bibinfo {author} {\bibfnamefont {A.}~\bibnamefont {Gily{\'e}n}}, \bibinfo {author} {\bibfnamefont {C.~T.}\ \bibnamefont {Hann}}, \bibinfo {author} {\bibfnamefont {M.~J.}\ \bibnamefont {Kastoryano}}, \bibinfo {author} {\bibfnamefont {E.~T.}\ \bibnamefont {Khabiboulline}}, \bibinfo {author} {\bibfnamefont {A.}~\bibnamefont {Kubica}},  \emph {et~al.},\ }\href {\doibase 10.48550/arXiv.2310.03011} {\bibfield  {journal} {\bibinfo  {journal} {arXiv preprint arXiv:2310.03011}\ } (\bibinfo {year} {2023}),\ 10.48550/arXiv.2310.03011}\BibitemShut {NoStop}%
\bibitem [{\citenamefont {Feynman}(2018)}]{feynman2018simulating}%
  \BibitemOpen
  \bibfield  {author} {\bibinfo {author} {\bibfnamefont {R.~P.}\ \bibnamefont {Feynman}},\ }in\ \href {\doibase https://doi.org/10.1007/BF02650179} {\emph {\bibinfo {booktitle} {Feynman and computation}}}\ (\bibinfo  {publisher} {cRc Press},\ \bibinfo {year} {2018})\ pp.\ \bibinfo {pages} {133--153}\BibitemShut {NoStop}%
\bibitem [{\citenamefont {Ponsioen}\ \emph {et~al.}(2019)\citenamefont {Ponsioen}, \citenamefont {Chung},\ and\ \citenamefont {Corboz}}]{ponsioen2019period}%
  \BibitemOpen
  \bibfield  {author} {\bibinfo {author} {\bibfnamefont {B.}~\bibnamefont {Ponsioen}}, \bibinfo {author} {\bibfnamefont {S.~S.}\ \bibnamefont {Chung}}, \ and\ \bibinfo {author} {\bibfnamefont {P.}~\bibnamefont {Corboz}},\ }\href {\doibase 10.1103/PhysRevB.100.195141} {\bibfield  {journal} {\bibinfo  {journal} {Physical Review B}\ }\textbf {\bibinfo {volume} {100}},\ \bibinfo {pages} {195141} (\bibinfo {year} {2019})}\BibitemShut {NoStop}%
\bibitem [{\citenamefont {Qin}\ \emph {et~al.}(2020)\citenamefont {Qin}, \citenamefont {Chung}, \citenamefont {Shi}, \citenamefont {Vitali}, \citenamefont {Hubig}, \citenamefont {Schollw{\"o}ck}, \citenamefont {White}, \citenamefont {Zhang},\ and\ \citenamefont {on~the Many-Electron~Problem)}}]{qin2020absence}%
  \BibitemOpen
  \bibfield  {author} {\bibinfo {author} {\bibfnamefont {M.}~\bibnamefont {Qin}}, \bibinfo {author} {\bibfnamefont {C.-M.}\ \bibnamefont {Chung}}, \bibinfo {author} {\bibfnamefont {H.}~\bibnamefont {Shi}}, \bibinfo {author} {\bibfnamefont {E.}~\bibnamefont {Vitali}}, \bibinfo {author} {\bibfnamefont {C.}~\bibnamefont {Hubig}}, \bibinfo {author} {\bibfnamefont {U.}~\bibnamefont {Schollw{\"o}ck}}, \bibinfo {author} {\bibfnamefont {S.~R.}\ \bibnamefont {White}}, \bibinfo {author} {\bibfnamefont {S.}~\bibnamefont {Zhang}}, \ and\ \bibinfo {author} {\bibfnamefont {S.~C.}\ \bibnamefont {on~the Many-Electron~Problem)}},\ }\href {\doibase 10.1103/PhysRevX.10.031016} {\bibfield  {journal} {\bibinfo  {journal} {Physical Review X}\ }\textbf {\bibinfo {volume} {10}},\ \bibinfo {pages} {031016} (\bibinfo {year} {2020})}\BibitemShut {NoStop}%
\bibitem [{\citenamefont {Xu}\ \emph {et~al.}(2024)\citenamefont {Xu}, \citenamefont {Chung}, \citenamefont {Qin}, \citenamefont {Schollw{\"o}ck}, \citenamefont {White},\ and\ \citenamefont {Zhang}}]{xu2024coexistence}%
  \BibitemOpen
  \bibfield  {author} {\bibinfo {author} {\bibfnamefont {H.}~\bibnamefont {Xu}}, \bibinfo {author} {\bibfnamefont {C.-M.}\ \bibnamefont {Chung}}, \bibinfo {author} {\bibfnamefont {M.}~\bibnamefont {Qin}}, \bibinfo {author} {\bibfnamefont {U.}~\bibnamefont {Schollw{\"o}ck}}, \bibinfo {author} {\bibfnamefont {S.~R.}\ \bibnamefont {White}}, \ and\ \bibinfo {author} {\bibfnamefont {S.}~\bibnamefont {Zhang}},\ }\href {\doibase 10.1126/science.adh7691} {\bibfield  {journal} {\bibinfo  {journal} {Science}\ }\textbf {\bibinfo {volume} {384}},\ \bibinfo {pages} {eadh7691} (\bibinfo {year} {2024})}\BibitemShut {NoStop}%
\bibitem [{\citenamefont {Haghshenas}\ \emph {et~al.}(2025)\citenamefont {Haghshenas}, \citenamefont {Chertkov}, \citenamefont {Mills}, \citenamefont {Kadow}, \citenamefont {Lin}, \citenamefont {Chen}, \citenamefont {Cade}, \citenamefont {Niesen}, \citenamefont {Begu{\v{s}}i{\'c}}, \citenamefont {Rudolph} \emph {et~al.}}]{haghshenas2025digital}%
  \BibitemOpen
  \bibfield  {author} {\bibinfo {author} {\bibfnamefont {R.}~\bibnamefont {Haghshenas}}, \bibinfo {author} {\bibfnamefont {E.}~\bibnamefont {Chertkov}}, \bibinfo {author} {\bibfnamefont {M.}~\bibnamefont {Mills}}, \bibinfo {author} {\bibfnamefont {W.}~\bibnamefont {Kadow}}, \bibinfo {author} {\bibfnamefont {S.-H.}\ \bibnamefont {Lin}}, \bibinfo {author} {\bibfnamefont {Y.-H.}\ \bibnamefont {Chen}}, \bibinfo {author} {\bibfnamefont {C.}~\bibnamefont {Cade}}, \bibinfo {author} {\bibfnamefont {I.}~\bibnamefont {Niesen}}, \bibinfo {author} {\bibfnamefont {T.}~\bibnamefont {Begu{\v{s}}i{\'c}}}, \bibinfo {author} {\bibfnamefont {M.~S.}\ \bibnamefont {Rudolph}},  \emph {et~al.},\ }\href {\doibase 10.48550/arXiv.2503.20870} {\bibfield  {journal} {\bibinfo  {journal} {arXiv preprint arXiv:2503.20870}\ } (\bibinfo {year} {2025}),\ 10.48550/arXiv.2503.20870}\BibitemShut {NoStop}%
\bibitem [{\citenamefont {Abanin}\ \emph {et~al.}(2025)\citenamefont {Abanin}, \citenamefont {Acharya}, \citenamefont {Aghababaie-Beni}, \citenamefont {Aigeldinger}, \citenamefont {Ajoy}, \citenamefont {Alcaraz}, \citenamefont {Aleiner}, \citenamefont {Andersen}, \citenamefont {Ansmann}, \citenamefont {Arute} \emph {et~al.}}]{abanin2025constructive}%
  \BibitemOpen
  \bibfield  {author} {\bibinfo {author} {\bibfnamefont {D.~A.}\ \bibnamefont {Abanin}}, \bibinfo {author} {\bibfnamefont {R.}~\bibnamefont {Acharya}}, \bibinfo {author} {\bibfnamefont {L.}~\bibnamefont {Aghababaie-Beni}}, \bibinfo {author} {\bibfnamefont {G.}~\bibnamefont {Aigeldinger}}, \bibinfo {author} {\bibfnamefont {A.}~\bibnamefont {Ajoy}}, \bibinfo {author} {\bibfnamefont {R.}~\bibnamefont {Alcaraz}}, \bibinfo {author} {\bibfnamefont {I.}~\bibnamefont {Aleiner}}, \bibinfo {author} {\bibfnamefont {T.~I.}\ \bibnamefont {Andersen}}, \bibinfo {author} {\bibfnamefont {M.}~\bibnamefont {Ansmann}}, \bibinfo {author} {\bibfnamefont {F.}~\bibnamefont {Arute}},  \emph {et~al.},\ }\href {\doibase 10.48550/arXiv.2506.10191} {\bibfield  {journal} {\bibinfo  {journal} {arXiv preprint arXiv:2506.10191}\ } (\bibinfo {year} {2025}),\ 10.48550/arXiv.2506.10191}\BibitemShut {NoStop}%
\bibitem [{\citenamefont {Andersen}\ \emph {et~al.}(2025)\citenamefont {Andersen}, \citenamefont {Astrakhantsev}, \citenamefont {Karamlou}, \citenamefont {Berndtsson}, \citenamefont {Motruk}, \citenamefont {Szasz}, \citenamefont {Gross}, \citenamefont {Schuckert}, \citenamefont {Westerhout}, \citenamefont {Zhang} \emph {et~al.}}]{andersen2025thermalization}%
  \BibitemOpen
  \bibfield  {author} {\bibinfo {author} {\bibfnamefont {T.~I.}\ \bibnamefont {Andersen}}, \bibinfo {author} {\bibfnamefont {N.}~\bibnamefont {Astrakhantsev}}, \bibinfo {author} {\bibfnamefont {A.~H.}\ \bibnamefont {Karamlou}}, \bibinfo {author} {\bibfnamefont {J.}~\bibnamefont {Berndtsson}}, \bibinfo {author} {\bibfnamefont {J.}~\bibnamefont {Motruk}}, \bibinfo {author} {\bibfnamefont {A.}~\bibnamefont {Szasz}}, \bibinfo {author} {\bibfnamefont {J.~A.}\ \bibnamefont {Gross}}, \bibinfo {author} {\bibfnamefont {A.}~\bibnamefont {Schuckert}}, \bibinfo {author} {\bibfnamefont {T.}~\bibnamefont {Westerhout}}, \bibinfo {author} {\bibfnamefont {Y.}~\bibnamefont {Zhang}},  \emph {et~al.},\ }\href {\doibase 10.48550/arXiv.2405.17385} {\bibfield  {journal} {\bibinfo  {journal} {Nature}\ }\textbf {\bibinfo {volume} {638}},\ \bibinfo {pages} {79} (\bibinfo {year} {2025})}\BibitemShut {NoStop}%
\bibitem [{\citenamefont {King}\ \emph {et~al.}(2025)\citenamefont {King}, \citenamefont {Nocera}, \citenamefont {Rams}, \citenamefont {Dziarmaga}, \citenamefont {Wiersema}, \citenamefont {Bernoudy}, \citenamefont {Raymond}, \citenamefont {Kaushal}, \citenamefont {Heinsdorf}, \citenamefont {Harris} \emph {et~al.}}]{king2025beyond}%
  \BibitemOpen
  \bibfield  {author} {\bibinfo {author} {\bibfnamefont {A.~D.}\ \bibnamefont {King}}, \bibinfo {author} {\bibfnamefont {A.}~\bibnamefont {Nocera}}, \bibinfo {author} {\bibfnamefont {M.~M.}\ \bibnamefont {Rams}}, \bibinfo {author} {\bibfnamefont {J.}~\bibnamefont {Dziarmaga}}, \bibinfo {author} {\bibfnamefont {R.}~\bibnamefont {Wiersema}}, \bibinfo {author} {\bibfnamefont {W.}~\bibnamefont {Bernoudy}}, \bibinfo {author} {\bibfnamefont {J.}~\bibnamefont {Raymond}}, \bibinfo {author} {\bibfnamefont {N.}~\bibnamefont {Kaushal}}, \bibinfo {author} {\bibfnamefont {N.}~\bibnamefont {Heinsdorf}}, \bibinfo {author} {\bibfnamefont {R.}~\bibnamefont {Harris}},  \emph {et~al.},\ }\href {\doibase 10.1126/science.ado6285} {\bibfield  {journal} {\bibinfo  {journal} {Science}\ }\textbf {\bibinfo {volume} {388}},\ \bibinfo {pages} {199} (\bibinfo {year} {2025})}\BibitemShut {NoStop}%
\bibitem [{\citenamefont {Kim}\ \emph {et~al.}(2023)\citenamefont {Kim}, \citenamefont {Eddins}, \citenamefont {Anand}, \citenamefont {Wei}, \citenamefont {Van Den~Berg}, \citenamefont {Rosenblatt}, \citenamefont {Nayfeh}, \citenamefont {Wu}, \citenamefont {Zaletel}, \citenamefont {Temme} \emph {et~al.}}]{kim2023evidence}%
  \BibitemOpen
  \bibfield  {author} {\bibinfo {author} {\bibfnamefont {Y.}~\bibnamefont {Kim}}, \bibinfo {author} {\bibfnamefont {A.}~\bibnamefont {Eddins}}, \bibinfo {author} {\bibfnamefont {S.}~\bibnamefont {Anand}}, \bibinfo {author} {\bibfnamefont {K.~X.}\ \bibnamefont {Wei}}, \bibinfo {author} {\bibfnamefont {E.}~\bibnamefont {Van Den~Berg}}, \bibinfo {author} {\bibfnamefont {S.}~\bibnamefont {Rosenblatt}}, \bibinfo {author} {\bibfnamefont {H.}~\bibnamefont {Nayfeh}}, \bibinfo {author} {\bibfnamefont {Y.}~\bibnamefont {Wu}}, \bibinfo {author} {\bibfnamefont {M.}~\bibnamefont {Zaletel}}, \bibinfo {author} {\bibfnamefont {K.}~\bibnamefont {Temme}},  \emph {et~al.},\ }\href {\doibase 10.1038/s41586-023-06096-3} {\bibfield  {journal} {\bibinfo  {journal} {Nature}\ }\textbf {\bibinfo {volume} {618}},\ \bibinfo {pages} {500} (\bibinfo {year} {2023})}\BibitemShut {NoStop}%
\bibitem [{\citenamefont {Granet}\ \emph {et~al.}(2025)\citenamefont {Granet}, \citenamefont {Lin}, \citenamefont {H{\'e}mery}, \citenamefont {Haghshenas}, \citenamefont {Andres-Martinez}, \citenamefont {Stephen}, \citenamefont {Ransford}, \citenamefont {Arkinstall}, \citenamefont {Allman}, \citenamefont {Campora} \emph {et~al.}}]{granet2025superconducting}%
  \BibitemOpen
  \bibfield  {author} {\bibinfo {author} {\bibfnamefont {E.}~\bibnamefont {Granet}}, \bibinfo {author} {\bibfnamefont {S.-H.}\ \bibnamefont {Lin}}, \bibinfo {author} {\bibfnamefont {K.}~\bibnamefont {H{\'e}mery}}, \bibinfo {author} {\bibfnamefont {R.}~\bibnamefont {Haghshenas}}, \bibinfo {author} {\bibfnamefont {P.}~\bibnamefont {Andres-Martinez}}, \bibinfo {author} {\bibfnamefont {D.~T.}\ \bibnamefont {Stephen}}, \bibinfo {author} {\bibfnamefont {A.}~\bibnamefont {Ransford}}, \bibinfo {author} {\bibfnamefont {J.}~\bibnamefont {Arkinstall}}, \bibinfo {author} {\bibfnamefont {M.}~\bibnamefont {Allman}}, \bibinfo {author} {\bibfnamefont {P.}~\bibnamefont {Campora}},  \emph {et~al.},\ }\href {\doibase 10.48550/arXiv.2511.02125} {\bibfield  {journal} {\bibinfo  {journal} {arXiv preprint arXiv:2511.02125}\ } (\bibinfo {year} {2025}),\ 10.48550/arXiv.2511.02125}\BibitemShut {NoStop}%
\bibitem [{\citenamefont {Foulkes}\ \emph {et~al.}(2001)\citenamefont {Foulkes}, \citenamefont {Mitas}, \citenamefont {Needs},\ and\ \citenamefont {Rajagopal}}]{foulkes2001quantum}%
  \BibitemOpen
  \bibfield  {author} {\bibinfo {author} {\bibfnamefont {W.~M.}\ \bibnamefont {Foulkes}}, \bibinfo {author} {\bibfnamefont {L.}~\bibnamefont {Mitas}}, \bibinfo {author} {\bibfnamefont {R.}~\bibnamefont {Needs}}, \ and\ \bibinfo {author} {\bibfnamefont {G.}~\bibnamefont {Rajagopal}},\ }\href {\doibase 10.1103/RevModPhys.73.33} {\bibfield  {journal} {\bibinfo  {journal} {Reviews of Modern Physics}\ }\textbf {\bibinfo {volume} {73}},\ \bibinfo {pages} {33} (\bibinfo {year} {2001})}\BibitemShut {NoStop}%
\bibitem [{\citenamefont {Albash}\ and\ \citenamefont {Lidar}(2018)}]{albash2018adiabatic}%
  \BibitemOpen
  \bibfield  {author} {\bibinfo {author} {\bibfnamefont {T.}~\bibnamefont {Albash}}\ and\ \bibinfo {author} {\bibfnamefont {D.~A.}\ \bibnamefont {Lidar}},\ }\href {\doibase 10.1103/RevModPhys.90.015002} {\bibfield  {journal} {\bibinfo  {journal} {Reviews of Modern Physics}\ }\textbf {\bibinfo {volume} {90}},\ \bibinfo {pages} {015002} (\bibinfo {year} {2018})}\BibitemShut {NoStop}%
\bibitem [{\citenamefont {Kitaev}(1995)}]{kitaev1995quantum}%
  \BibitemOpen
  \bibfield  {author} {\bibinfo {author} {\bibfnamefont {A.~Y.}\ \bibnamefont {Kitaev}},\ }\href {\doibase 10.48550/arXiv.quant-ph/9511026} {\bibfield  {journal} {\bibinfo  {journal} {arXiv preprint quant-ph/9511026}\ } (\bibinfo {year} {1995}),\ 10.48550/arXiv.quant-ph/9511026}\BibitemShut {NoStop}%
\bibitem [{\citenamefont {Chowdhury}\ and\ \citenamefont {Somma}(2016)}]{chowdhury2016quantum}%
  \BibitemOpen
  \bibfield  {author} {\bibinfo {author} {\bibfnamefont {A.~N.}\ \bibnamefont {Chowdhury}}\ and\ \bibinfo {author} {\bibfnamefont {R.~D.}\ \bibnamefont {Somma}},\ }\href {\doibase 10.48550/arXiv.1603.02940} {\bibfield  {journal} {\bibinfo  {journal} {arXiv preprint arXiv:1603.02940}\ } (\bibinfo {year} {2016}),\ 10.48550/arXiv.1603.02940}\BibitemShut {NoStop}%
\bibitem [{\citenamefont {Motta}\ \emph {et~al.}(2020)\citenamefont {Motta}, \citenamefont {Sun}, \citenamefont {Tan}, \citenamefont {O’Rourke}, \citenamefont {Ye}, \citenamefont {Minnich}, \citenamefont {Brandao},\ and\ \citenamefont {Chan}}]{motta2020determining}%
  \BibitemOpen
  \bibfield  {author} {\bibinfo {author} {\bibfnamefont {M.}~\bibnamefont {Motta}}, \bibinfo {author} {\bibfnamefont {C.}~\bibnamefont {Sun}}, \bibinfo {author} {\bibfnamefont {A.~T.}\ \bibnamefont {Tan}}, \bibinfo {author} {\bibfnamefont {M.~J.}\ \bibnamefont {O’Rourke}}, \bibinfo {author} {\bibfnamefont {E.}~\bibnamefont {Ye}}, \bibinfo {author} {\bibfnamefont {A.~J.}\ \bibnamefont {Minnich}}, \bibinfo {author} {\bibfnamefont {F.~G.}\ \bibnamefont {Brandao}}, \ and\ \bibinfo {author} {\bibfnamefont {G.~K.-L.}\ \bibnamefont {Chan}},\ }\href {\doibase 10.1038/s41567-019-0704-4} {\bibfield  {journal} {\bibinfo  {journal} {Nature Physics}\ }\textbf {\bibinfo {volume} {16}},\ \bibinfo {pages} {205} (\bibinfo {year} {2020})}\BibitemShut {NoStop}%
\bibitem [{\citenamefont {Coopmans}\ \emph {et~al.}(2023)\citenamefont {Coopmans}, \citenamefont {Kikuchi},\ and\ \citenamefont {Benedetti}}]{coopmans2023predicting}%
  \BibitemOpen
  \bibfield  {author} {\bibinfo {author} {\bibfnamefont {L.}~\bibnamefont {Coopmans}}, \bibinfo {author} {\bibfnamefont {Y.}~\bibnamefont {Kikuchi}}, \ and\ \bibinfo {author} {\bibfnamefont {M.}~\bibnamefont {Benedetti}},\ }\href {\doibase 10.1103/PRXQuantum.4.010305} {\bibfield  {journal} {\bibinfo  {journal} {PRX Quantum}\ }\textbf {\bibinfo {volume} {4}},\ \bibinfo {pages} {010305} (\bibinfo {year} {2023})}\BibitemShut {NoStop}%
\bibitem [{\citenamefont {Shtanko}\ and\ \citenamefont {Movassagh}(2021)}]{shtanko2021preparing}%
  \BibitemOpen
  \bibfield  {author} {\bibinfo {author} {\bibfnamefont {O.}~\bibnamefont {Shtanko}}\ and\ \bibinfo {author} {\bibfnamefont {R.}~\bibnamefont {Movassagh}},\ }\href {\doibase 10.48550/arXiv.2112.14688} {\bibfield  {journal} {\bibinfo  {journal} {arXiv preprint arXiv:2112.14688}\ } (\bibinfo {year} {2021}),\ 10.48550/arXiv.2112.14688}\BibitemShut {NoStop}%
\bibitem [{\citenamefont {Temme}\ \emph {et~al.}(2011)\citenamefont {Temme}, \citenamefont {Osborne}, \citenamefont {Vollbrecht}, \citenamefont {Poulin},\ and\ \citenamefont {Verstraete}}]{temme2011quantum}%
  \BibitemOpen
  \bibfield  {author} {\bibinfo {author} {\bibfnamefont {K.}~\bibnamefont {Temme}}, \bibinfo {author} {\bibfnamefont {T.~J.}\ \bibnamefont {Osborne}}, \bibinfo {author} {\bibfnamefont {K.~G.}\ \bibnamefont {Vollbrecht}}, \bibinfo {author} {\bibfnamefont {D.}~\bibnamefont {Poulin}}, \ and\ \bibinfo {author} {\bibfnamefont {F.}~\bibnamefont {Verstraete}},\ }\href {\doibase 10.1038/nature09770} {\bibfield  {journal} {\bibinfo  {journal} {Nature}\ }\textbf {\bibinfo {volume} {471}},\ \bibinfo {pages} {87} (\bibinfo {year} {2011})}\BibitemShut {NoStop}%
\bibitem [{\citenamefont {Cohn}\ \emph {et~al.}(2020)\citenamefont {Cohn}, \citenamefont {Yang}, \citenamefont {Najafi}, \citenamefont {Jones},\ and\ \citenamefont {Freericks}}]{cohn2020minimal}%
  \BibitemOpen
  \bibfield  {author} {\bibinfo {author} {\bibfnamefont {J.}~\bibnamefont {Cohn}}, \bibinfo {author} {\bibfnamefont {F.}~\bibnamefont {Yang}}, \bibinfo {author} {\bibfnamefont {K.}~\bibnamefont {Najafi}}, \bibinfo {author} {\bibfnamefont {B.}~\bibnamefont {Jones}}, \ and\ \bibinfo {author} {\bibfnamefont {J.~K.}\ \bibnamefont {Freericks}},\ }\href {\doibase 10.1103/PhysRevA.102.022622} {\bibfield  {journal} {\bibinfo  {journal} {Physical Review A}\ }\textbf {\bibinfo {volume} {102}},\ \bibinfo {pages} {022622} (\bibinfo {year} {2020})}\BibitemShut {NoStop}%
\bibitem [{\citenamefont {Chen}\ \emph {et~al.}(2023{\natexlab{a}})\citenamefont {Chen}, \citenamefont {Kastoryano},\ and\ \citenamefont {Gily{\'e}n}}]{chen2023efficient}%
  \BibitemOpen
  \bibfield  {author} {\bibinfo {author} {\bibfnamefont {C.-F.}\ \bibnamefont {Chen}}, \bibinfo {author} {\bibfnamefont {M.~J.}\ \bibnamefont {Kastoryano}}, \ and\ \bibinfo {author} {\bibfnamefont {A.}~\bibnamefont {Gily{\'e}n}},\ }\href {\doibase 10.48550/arXiv.2311.09207} {\bibfield  {journal} {\bibinfo  {journal} {arXiv preprint arXiv:2311.09207}\ } (\bibinfo {year} {2023}{\natexlab{a}}),\ 10.48550/arXiv.2311.09207}\BibitemShut {NoStop}%
\bibitem [{\citenamefont {Granet}\ and\ \citenamefont {Dreyer}(2024)}]{granet2024noise}%
  \BibitemOpen
  \bibfield  {author} {\bibinfo {author} {\bibfnamefont {E.}~\bibnamefont {Granet}}\ and\ \bibinfo {author} {\bibfnamefont {H.}~\bibnamefont {Dreyer}},\ }\href {\doibase 10.48550/arXiv.2401.02207} {\bibfield  {journal} {\bibinfo  {journal} {arXiv preprint arXiv:2401.02207}\ } (\bibinfo {year} {2024}),\ 10.48550/arXiv.2401.02207}\BibitemShut {NoStop}%
\bibitem [{\citenamefont {Bergamaschi}\ \emph {et~al.}(2024)\citenamefont {Bergamaschi}, \citenamefont {Chen},\ and\ \citenamefont {Liu}}]{bergamaschi2024quantum}%
  \BibitemOpen
  \bibfield  {author} {\bibinfo {author} {\bibfnamefont {T.}~\bibnamefont {Bergamaschi}}, \bibinfo {author} {\bibfnamefont {C.-F.}\ \bibnamefont {Chen}}, \ and\ \bibinfo {author} {\bibfnamefont {Y.}~\bibnamefont {Liu}},\ }in\ \href {\doibase 10.1109/FOCS61266.2024.00071} {\emph {\bibinfo {booktitle} {2024 IEEE 65th Annual Symposium on Foundations of Computer Science (FOCS)}}}\ (\bibinfo {organization} {IEEE},\ \bibinfo {year} {2024})\ pp.\ \bibinfo {pages} {1063--1085}\BibitemShut {NoStop}%
\bibitem [{\citenamefont {Chen}\ \emph {et~al.}(2023{\natexlab{b}})\citenamefont {Chen}, \citenamefont {Kastoryano}, \citenamefont {Brand{\~a}o},\ and\ \citenamefont {Gily{\'e}n}}]{chen2023quantum}%
  \BibitemOpen
  \bibfield  {author} {\bibinfo {author} {\bibfnamefont {C.-F.}\ \bibnamefont {Chen}}, \bibinfo {author} {\bibfnamefont {M.~J.}\ \bibnamefont {Kastoryano}}, \bibinfo {author} {\bibfnamefont {F.~G.}\ \bibnamefont {Brand{\~a}o}}, \ and\ \bibinfo {author} {\bibfnamefont {A.}~\bibnamefont {Gily{\'e}n}},\ }\href {\doibase 10.48550/arXiv.2303.18224} {\bibfield  {journal} {\bibinfo  {journal} {arXiv preprint arXiv:2303.18224}\ } (\bibinfo {year} {2023}{\natexlab{b}}),\ 10.48550/arXiv.2303.18224}\BibitemShut {NoStop}%
\bibitem [{\citenamefont {Brunner}\ \emph {et~al.}(2024)\citenamefont {Brunner}, \citenamefont {Coopmans}, \citenamefont {Matos}, \citenamefont {Rosenkranz}, \citenamefont {Sauvage},\ and\ \citenamefont {Kikuchi}}]{brunner2024lindblad}%
  \BibitemOpen
  \bibfield  {author} {\bibinfo {author} {\bibfnamefont {E.}~\bibnamefont {Brunner}}, \bibinfo {author} {\bibfnamefont {L.}~\bibnamefont {Coopmans}}, \bibinfo {author} {\bibfnamefont {G.}~\bibnamefont {Matos}}, \bibinfo {author} {\bibfnamefont {M.}~\bibnamefont {Rosenkranz}}, \bibinfo {author} {\bibfnamefont {F.}~\bibnamefont {Sauvage}}, \ and\ \bibinfo {author} {\bibfnamefont {Y.}~\bibnamefont {Kikuchi}},\ }\href {\doibase 10.48550/arXiv.2412.17706} {\bibfield  {journal} {\bibinfo  {journal} {arXiv preprint arXiv:2412.17706}\ } (\bibinfo {year} {2024}),\ 10.48550/arXiv.2412.17706}\BibitemShut {NoStop}%
\bibitem [{\citenamefont {Rall}\ \emph {et~al.}(2023)\citenamefont {Rall}, \citenamefont {Wang},\ and\ \citenamefont {Wocjan}}]{rall2023thermal}%
  \BibitemOpen
  \bibfield  {author} {\bibinfo {author} {\bibfnamefont {P.}~\bibnamefont {Rall}}, \bibinfo {author} {\bibfnamefont {C.}~\bibnamefont {Wang}}, \ and\ \bibinfo {author} {\bibfnamefont {P.}~\bibnamefont {Wocjan}},\ }\href {\doibase 10.22331/q-2023-10-10-1132} {\bibfield  {journal} {\bibinfo  {journal} {Quantum}\ }\textbf {\bibinfo {volume} {7}},\ \bibinfo {pages} {1132} (\bibinfo {year} {2023})}\BibitemShut {NoStop}%
\bibitem [{\citenamefont {Haug}\ and\ \citenamefont {Bharti}(2022)}]{haug2022generalized}%
  \BibitemOpen
  \bibfield  {author} {\bibinfo {author} {\bibfnamefont {T.}~\bibnamefont {Haug}}\ and\ \bibinfo {author} {\bibfnamefont {K.}~\bibnamefont {Bharti}},\ }\href {\doibase 10.1088/2058-9565/ac83e7} {\bibfield  {journal} {\bibinfo  {journal} {Quantum Science \& Technology}\ }\textbf {\bibinfo {volume} {7}},\ \bibinfo {pages} {045019} (\bibinfo {year} {2022})}\BibitemShut {NoStop}%
\bibitem [{\citenamefont {Consiglio}\ \emph {et~al.}(2024)\citenamefont {Consiglio}, \citenamefont {Settino}, \citenamefont {Giordano}, \citenamefont {Mastroianni}, \citenamefont {Plastina}, \citenamefont {Lorenzo}, \citenamefont {Maniscalco}, \citenamefont {Goold},\ and\ \citenamefont {Apollaro}}]{consiglio2024variational}%
  \BibitemOpen
  \bibfield  {author} {\bibinfo {author} {\bibfnamefont {M.}~\bibnamefont {Consiglio}}, \bibinfo {author} {\bibfnamefont {J.}~\bibnamefont {Settino}}, \bibinfo {author} {\bibfnamefont {A.}~\bibnamefont {Giordano}}, \bibinfo {author} {\bibfnamefont {C.}~\bibnamefont {Mastroianni}}, \bibinfo {author} {\bibfnamefont {F.}~\bibnamefont {Plastina}}, \bibinfo {author} {\bibfnamefont {S.}~\bibnamefont {Lorenzo}}, \bibinfo {author} {\bibfnamefont {S.}~\bibnamefont {Maniscalco}}, \bibinfo {author} {\bibfnamefont {J.}~\bibnamefont {Goold}}, \ and\ \bibinfo {author} {\bibfnamefont {T.~J.}\ \bibnamefont {Apollaro}},\ }\href {\doibase 10.1103/PhysRevA.110.012445} {\bibfield  {journal} {\bibinfo  {journal} {Physical Review A}\ }\textbf {\bibinfo {volume} {110}},\ \bibinfo {pages} {012445} (\bibinfo {year} {2024})}\BibitemShut {NoStop}%
\bibitem [{\citenamefont {Sewell}\ \emph {et~al.}(2022)\citenamefont {Sewell}, \citenamefont {White},\ and\ \citenamefont {Swingle}}]{sewell2022thermal}%
  \BibitemOpen
  \bibfield  {author} {\bibinfo {author} {\bibfnamefont {T.~J.}\ \bibnamefont {Sewell}}, \bibinfo {author} {\bibfnamefont {C.~D.}\ \bibnamefont {White}}, \ and\ \bibinfo {author} {\bibfnamefont {B.}~\bibnamefont {Swingle}},\ }\href {\doibase 10.48550/arXiv.2210.16419} {\bibfield  {journal} {\bibinfo  {journal} {arXiv preprint arXiv:2210.16419}\ } (\bibinfo {year} {2022}),\ 10.48550/arXiv.2210.16419}\BibitemShut {NoStop}%
\bibitem [{\citenamefont {Martyn}\ and\ \citenamefont {Swingle}(2019)}]{martyn2019product}%
  \BibitemOpen
  \bibfield  {author} {\bibinfo {author} {\bibfnamefont {J.}~\bibnamefont {Martyn}}\ and\ \bibinfo {author} {\bibfnamefont {B.}~\bibnamefont {Swingle}},\ }\href {\doibase 10.1103/PhysRevA.100.032107} {\bibfield  {journal} {\bibinfo  {journal} {Physical Review A}\ }\textbf {\bibinfo {volume} {100}},\ \bibinfo {pages} {032107} (\bibinfo {year} {2019})}\BibitemShut {NoStop}%
\bibitem [{\citenamefont {Wu}\ and\ \citenamefont {Hsieh}(2019)}]{wu2019variational}%
  \BibitemOpen
  \bibfield  {author} {\bibinfo {author} {\bibfnamefont {J.}~\bibnamefont {Wu}}\ and\ \bibinfo {author} {\bibfnamefont {T.~H.}\ \bibnamefont {Hsieh}},\ }\href {\doibase 10.1103/PhysRevLett.123.220502} {\bibfield  {journal} {\bibinfo  {journal} {Physical review letters}\ }\textbf {\bibinfo {volume} {123}},\ \bibinfo {pages} {220502} (\bibinfo {year} {2019})}\BibitemShut {NoStop}%
\bibitem [{\citenamefont {Chowdhury}\ \emph {et~al.}(2020)\citenamefont {Chowdhury}, \citenamefont {Low},\ and\ \citenamefont {Wiebe}}]{chowdhury2020variational}%
  \BibitemOpen
  \bibfield  {author} {\bibinfo {author} {\bibfnamefont {A.~N.}\ \bibnamefont {Chowdhury}}, \bibinfo {author} {\bibfnamefont {G.~H.}\ \bibnamefont {Low}}, \ and\ \bibinfo {author} {\bibfnamefont {N.}~\bibnamefont {Wiebe}},\ }\href {\doibase 10.48550/arXiv.2002.00055} {\bibfield  {journal} {\bibinfo  {journal} {arXiv preprint arXiv:2002.00055}\ } (\bibinfo {year} {2020}),\ 10.48550/arXiv.2002.00055}\BibitemShut {NoStop}%
\bibitem [{\citenamefont {Wang}\ \emph {et~al.}(2021)\citenamefont {Wang}, \citenamefont {Li},\ and\ \citenamefont {Wang}}]{wang2021variational}%
  \BibitemOpen
  \bibfield  {author} {\bibinfo {author} {\bibfnamefont {Y.}~\bibnamefont {Wang}}, \bibinfo {author} {\bibfnamefont {G.}~\bibnamefont {Li}}, \ and\ \bibinfo {author} {\bibfnamefont {X.}~\bibnamefont {Wang}},\ }\href {\doibase 10.1103/PhysRevApplied.16.054035} {\bibfield  {journal} {\bibinfo  {journal} {Physical Review Applied}\ }\textbf {\bibinfo {volume} {16}},\ \bibinfo {pages} {054035} (\bibinfo {year} {2021})}\BibitemShut {NoStop}%
\bibitem [{\citenamefont {Warren}\ \emph {et~al.}(2022)\citenamefont {Warren}, \citenamefont {Zhu}, \citenamefont {Mayhall}, \citenamefont {Barnes},\ and\ \citenamefont {Economou}}]{warren2022adaptive}%
  \BibitemOpen
  \bibfield  {author} {\bibinfo {author} {\bibfnamefont {A.}~\bibnamefont {Warren}}, \bibinfo {author} {\bibfnamefont {L.}~\bibnamefont {Zhu}}, \bibinfo {author} {\bibfnamefont {N.~J.}\ \bibnamefont {Mayhall}}, \bibinfo {author} {\bibfnamefont {E.}~\bibnamefont {Barnes}}, \ and\ \bibinfo {author} {\bibfnamefont {S.~E.}\ \bibnamefont {Economou}},\ }\href {\doibase 10.48550/arXiv.2203.12757} {\bibfield  {journal} {\bibinfo  {journal} {arXiv preprint arXiv:2203.12757}\ } (\bibinfo {year} {2022}),\ 10.48550/arXiv.2203.12757}\BibitemShut {NoStop}%
\bibitem [{\citenamefont {Foldager}\ \emph {et~al.}(2022)\citenamefont {Foldager}, \citenamefont {Pesah},\ and\ \citenamefont {Hansen}}]{foldager2022noise}%
  \BibitemOpen
  \bibfield  {author} {\bibinfo {author} {\bibfnamefont {J.}~\bibnamefont {Foldager}}, \bibinfo {author} {\bibfnamefont {A.}~\bibnamefont {Pesah}}, \ and\ \bibinfo {author} {\bibfnamefont {L.~K.}\ \bibnamefont {Hansen}},\ }\href {\doibase 10.48550/arXiv.2111.03935} {\bibfield  {journal} {\bibinfo  {journal} {Scientific reports}\ }\textbf {\bibinfo {volume} {12}},\ \bibinfo {pages} {3862} (\bibinfo {year} {2022})}\BibitemShut {NoStop}%
\bibitem [{\citenamefont {Lu}\ \emph {et~al.}(2021)\citenamefont {Lu}, \citenamefont {Ba{\~n}uls},\ and\ \citenamefont {Cirac}}]{lu2021algorithms}%
  \BibitemOpen
  \bibfield  {author} {\bibinfo {author} {\bibfnamefont {S.}~\bibnamefont {Lu}}, \bibinfo {author} {\bibfnamefont {M.~C.}\ \bibnamefont {Ba{\~n}uls}}, \ and\ \bibinfo {author} {\bibfnamefont {J.~I.}\ \bibnamefont {Cirac}},\ }\href {\doibase 10.1103/PRXQuantum.2.020321} {\bibfield  {journal} {\bibinfo  {journal} {PRX quantum}\ }\textbf {\bibinfo {volume} {2}},\ \bibinfo {pages} {020321} (\bibinfo {year} {2021})}\BibitemShut {NoStop}%
\bibitem [{\citenamefont {Holmes}\ \emph {et~al.}(2022)\citenamefont {Holmes}, \citenamefont {Muraleedharan}, \citenamefont {Somma}, \citenamefont {Subasi},\ and\ \citenamefont {{\c{S}}ahino{\u{g}}lu}}]{holmes2022quantum}%
  \BibitemOpen
  \bibfield  {author} {\bibinfo {author} {\bibfnamefont {Z.}~\bibnamefont {Holmes}}, \bibinfo {author} {\bibfnamefont {G.}~\bibnamefont {Muraleedharan}}, \bibinfo {author} {\bibfnamefont {R.~D.}\ \bibnamefont {Somma}}, \bibinfo {author} {\bibfnamefont {Y.}~\bibnamefont {Subasi}}, \ and\ \bibinfo {author} {\bibfnamefont {B.}~\bibnamefont {{\c{S}}ahino{\u{g}}lu}},\ }\href {\doibase 10.22331/q-2022-10-06-825} {\bibfield  {journal} {\bibinfo  {journal} {Quantum}\ }\textbf {\bibinfo {volume} {6}},\ \bibinfo {pages} {825} (\bibinfo {year} {2022})}\BibitemShut {NoStop}%
\bibitem [{\citenamefont {Poulin}\ and\ \citenamefont {Wocjan}(2009)}]{poulin2009sampling}%
  \BibitemOpen
  \bibfield  {author} {\bibinfo {author} {\bibfnamefont {D.}~\bibnamefont {Poulin}}\ and\ \bibinfo {author} {\bibfnamefont {P.}~\bibnamefont {Wocjan}},\ }\href {\doibase 10.1103/PhysRevLett.103.220502} {\bibfield  {journal} {\bibinfo  {journal} {Physical review letters}\ }\textbf {\bibinfo {volume} {103}},\ \bibinfo {pages} {220502} (\bibinfo {year} {2009})}\BibitemShut {NoStop}%
\bibitem [{\citenamefont {Terhal}\ and\ \citenamefont {DiVincenzo}(2000)}]{terhal2000problem}%
  \BibitemOpen
  \bibfield  {author} {\bibinfo {author} {\bibfnamefont {B.~M.}\ \bibnamefont {Terhal}}\ and\ \bibinfo {author} {\bibfnamefont {D.~P.}\ \bibnamefont {DiVincenzo}},\ }\href {\doibase 10.1103/PhysRevA.61.022301} {\bibfield  {journal} {\bibinfo  {journal} {Physical Review A}\ }\textbf {\bibinfo {volume} {61}},\ \bibinfo {pages} {022301} (\bibinfo {year} {2000})}\BibitemShut {NoStop}%
\bibitem [{\citenamefont {Riera}\ \emph {et~al.}(2012)\citenamefont {Riera}, \citenamefont {Gogolin},\ and\ \citenamefont {Eisert}}]{riera2012thermalization}%
  \BibitemOpen
  \bibfield  {author} {\bibinfo {author} {\bibfnamefont {A.}~\bibnamefont {Riera}}, \bibinfo {author} {\bibfnamefont {C.}~\bibnamefont {Gogolin}}, \ and\ \bibinfo {author} {\bibfnamefont {J.}~\bibnamefont {Eisert}},\ }\href {\doibase 10.1103/PhysRevLett.108.080402} {\bibfield  {journal} {\bibinfo  {journal} {Physical review letters}\ }\textbf {\bibinfo {volume} {108}},\ \bibinfo {pages} {080402} (\bibinfo {year} {2012})}\BibitemShut {NoStop}%
\bibitem [{\citenamefont {Bilgin}\ and\ \citenamefont {Boixo}(2010)}]{bilgin2010preparing}%
  \BibitemOpen
  \bibfield  {author} {\bibinfo {author} {\bibfnamefont {E.}~\bibnamefont {Bilgin}}\ and\ \bibinfo {author} {\bibfnamefont {S.}~\bibnamefont {Boixo}},\ }\href {\doibase 10.1103/PhysRevLett.105.170405} {\bibfield  {journal} {\bibinfo  {journal} {Physical review letters}\ }\textbf {\bibinfo {volume} {105}},\ \bibinfo {pages} {170405} (\bibinfo {year} {2010})}\BibitemShut {NoStop}%
\bibitem [{\citenamefont {H{\'e}mery}\ \emph {et~al.}(2024)\citenamefont {H{\'e}mery}, \citenamefont {Ghanem}, \citenamefont {Crane}, \citenamefont {Campbell}, \citenamefont {Dreiling}, \citenamefont {Figgatt}, \citenamefont {Foltz}, \citenamefont {Gaebler}, \citenamefont {Johansen}, \citenamefont {Mills} \emph {et~al.}}]{hemery2024measuring}%
  \BibitemOpen
  \bibfield  {author} {\bibinfo {author} {\bibfnamefont {K.}~\bibnamefont {H{\'e}mery}}, \bibinfo {author} {\bibfnamefont {K.}~\bibnamefont {Ghanem}}, \bibinfo {author} {\bibfnamefont {E.}~\bibnamefont {Crane}}, \bibinfo {author} {\bibfnamefont {S.~L.}\ \bibnamefont {Campbell}}, \bibinfo {author} {\bibfnamefont {J.~M.}\ \bibnamefont {Dreiling}}, \bibinfo {author} {\bibfnamefont {C.}~\bibnamefont {Figgatt}}, \bibinfo {author} {\bibfnamefont {C.}~\bibnamefont {Foltz}}, \bibinfo {author} {\bibfnamefont {J.~P.}\ \bibnamefont {Gaebler}}, \bibinfo {author} {\bibfnamefont {J.}~\bibnamefont {Johansen}}, \bibinfo {author} {\bibfnamefont {M.}~\bibnamefont {Mills}},  \emph {et~al.},\ }\href {\doibase 10.1103/PRXQuantum.5.030323} {\bibfield  {journal} {\bibinfo  {journal} {PRX Quantum}\ }\textbf {\bibinfo {volume} {5}},\ \bibinfo {pages} {030323} (\bibinfo {year} {2024})}\BibitemShut {NoStop}%
\bibitem [{\citenamefont {Mori}\ \emph {et~al.}(2018)\citenamefont {Mori}, \citenamefont {Ikeda}, \citenamefont {Kaminishi},\ and\ \citenamefont {Ueda}}]{mori2018thermalization}%
  \BibitemOpen
  \bibfield  {author} {\bibinfo {author} {\bibfnamefont {T.}~\bibnamefont {Mori}}, \bibinfo {author} {\bibfnamefont {T.~N.}\ \bibnamefont {Ikeda}}, \bibinfo {author} {\bibfnamefont {E.}~\bibnamefont {Kaminishi}}, \ and\ \bibinfo {author} {\bibfnamefont {M.}~\bibnamefont {Ueda}},\ }\href {\doibase 10.1088/1361-6455/aabcdf} {\bibfield  {journal} {\bibinfo  {journal} {Journal of Physics B: Atomic, Molecular and Optical Physics}\ }\textbf {\bibinfo {volume} {51}},\ \bibinfo {pages} {112001} (\bibinfo {year} {2018})}\BibitemShut {NoStop}%
\bibitem [{\citenamefont {Yarloo}\ \emph {et~al.}(2024)\citenamefont {Yarloo}, \citenamefont {Zhang},\ and\ \citenamefont {Nielsen}}]{yarloo2024adiabatic}%
  \BibitemOpen
  \bibfield  {author} {\bibinfo {author} {\bibfnamefont {H.}~\bibnamefont {Yarloo}}, \bibinfo {author} {\bibfnamefont {H.-C.}\ \bibnamefont {Zhang}}, \ and\ \bibinfo {author} {\bibfnamefont {A.~E.}\ \bibnamefont {Nielsen}},\ }\href {\doibase 10.1103/PRXQuantum.5.020365} {\bibfield  {journal} {\bibinfo  {journal} {PRX Quantum}\ }\textbf {\bibinfo {volume} {5}},\ \bibinfo {pages} {020365} (\bibinfo {year} {2024})}\BibitemShut {NoStop}%
\bibitem [{\citenamefont {Greenblatt}\ \emph {et~al.}(2024)\citenamefont {Greenblatt}, \citenamefont {Lange}, \citenamefont {Marcelli},\ and\ \citenamefont {Porta}}]{greenblatt2024adiabatic}%
  \BibitemOpen
  \bibfield  {author} {\bibinfo {author} {\bibfnamefont {R.~L.}\ \bibnamefont {Greenblatt}}, \bibinfo {author} {\bibfnamefont {M.}~\bibnamefont {Lange}}, \bibinfo {author} {\bibfnamefont {G.}~\bibnamefont {Marcelli}}, \ and\ \bibinfo {author} {\bibfnamefont {M.}~\bibnamefont {Porta}},\ }\href {\doibase 10.1007/s00220-023-04903-6} {\bibfield  {journal} {\bibinfo  {journal} {Communications in Mathematical Physics}\ }\textbf {\bibinfo {volume} {405}},\ \bibinfo {pages} {75} (\bibinfo {year} {2024})}\BibitemShut {NoStop}%
\bibitem [{\citenamefont {Il~‘in}\ \emph {et~al.}(2021)\citenamefont {Il~‘in}, \citenamefont {Aristova},\ and\ \citenamefont {Lychkovskiy}}]{il2021adiabatic}%
  \BibitemOpen
  \bibfield  {author} {\bibinfo {author} {\bibfnamefont {N.}~\bibnamefont {Il~‘in}}, \bibinfo {author} {\bibfnamefont {A.}~\bibnamefont {Aristova}}, \ and\ \bibinfo {author} {\bibfnamefont {O.}~\bibnamefont {Lychkovskiy}},\ }\href {\doibase 10.1103/PhysRevA.104.L030202} {\bibfield  {journal} {\bibinfo  {journal} {Physical Review A}\ }\textbf {\bibinfo {volume} {104}},\ \bibinfo {pages} {L030202} (\bibinfo {year} {2021})}\BibitemShut {NoStop}%
\bibitem [{\citenamefont {Zuo}\ \emph {et~al.}(2024)\citenamefont {Zuo}, \citenamefont {Yang}, \citenamefont {Liu},\ and\ \citenamefont {Liu}}]{zuo2024work}%
  \BibitemOpen
  \bibfield  {author} {\bibinfo {author} {\bibfnamefont {Y.}~\bibnamefont {Zuo}}, \bibinfo {author} {\bibfnamefont {Q.}~\bibnamefont {Yang}}, \bibinfo {author} {\bibfnamefont {B.-G.}\ \bibnamefont {Liu}}, \ and\ \bibinfo {author} {\bibfnamefont {D.~E.}\ \bibnamefont {Liu}},\ }\href {\doibase 10.1103/PhysRevE.110.L022105} {\bibfield  {journal} {\bibinfo  {journal} {Physical Review E}\ }\textbf {\bibinfo {volume} {110}},\ \bibinfo {pages} {L022105} (\bibinfo {year} {2024})}\BibitemShut {NoStop}%
\bibitem [{\citenamefont {Irmejs}\ \emph {et~al.}(2025)\citenamefont {Irmejs}, \citenamefont {Ba{\~n}uls},\ and\ \citenamefont {Cirac}}]{irmejs2025quasi}%
  \BibitemOpen
  \bibfield  {author} {\bibinfo {author} {\bibfnamefont {R.}~\bibnamefont {Irmejs}}, \bibinfo {author} {\bibfnamefont {M.~C.}\ \bibnamefont {Ba{\~n}uls}}, \ and\ \bibinfo {author} {\bibfnamefont {J.~I.}\ \bibnamefont {Cirac}},\ }\href {\doibase 10.48550/arXiv.2505.20042} {\bibfield  {journal} {\bibinfo  {journal} {arXiv preprint arXiv:2505.20042}\ } (\bibinfo {year} {2025}),\ 10.48550/arXiv.2505.20042}\BibitemShut {NoStop}%
\bibitem [{\citenamefont {Plastina}\ \emph {et~al.}(2014)\citenamefont {Plastina}, \citenamefont {Alecce}, \citenamefont {Apollaro}, \citenamefont {Falcone}, \citenamefont {Francica}, \citenamefont {Galve}, \citenamefont {Lo~Gullo},\ and\ \citenamefont {Zambrini}}]{plastina2014irreversible}%
  \BibitemOpen
  \bibfield  {author} {\bibinfo {author} {\bibfnamefont {F.}~\bibnamefont {Plastina}}, \bibinfo {author} {\bibfnamefont {A.}~\bibnamefont {Alecce}}, \bibinfo {author} {\bibfnamefont {T.~J.}\ \bibnamefont {Apollaro}}, \bibinfo {author} {\bibfnamefont {G.}~\bibnamefont {Falcone}}, \bibinfo {author} {\bibfnamefont {G.}~\bibnamefont {Francica}}, \bibinfo {author} {\bibfnamefont {F.}~\bibnamefont {Galve}}, \bibinfo {author} {\bibfnamefont {N.}~\bibnamefont {Lo~Gullo}}, \ and\ \bibinfo {author} {\bibfnamefont {R.}~\bibnamefont {Zambrini}},\ }\href {\doibase 10.1103/PhysRevLett.113.260601} {\bibfield  {journal} {\bibinfo  {journal} {Physical review letters}\ }\textbf {\bibinfo {volume} {113}},\ \bibinfo {pages} {260601} (\bibinfo {year} {2014})}\BibitemShut {NoStop}%
\bibitem [{\citenamefont {Francica}\ \emph {et~al.}(2019)\citenamefont {Francica}, \citenamefont {Goold},\ and\ \citenamefont {Plastina}}]{francica2019role}%
  \BibitemOpen
  \bibfield  {author} {\bibinfo {author} {\bibfnamefont {G.}~\bibnamefont {Francica}}, \bibinfo {author} {\bibfnamefont {J.}~\bibnamefont {Goold}}, \ and\ \bibinfo {author} {\bibfnamefont {F.}~\bibnamefont {Plastina}},\ }\href {\doibase 10.1103/PhysRevE.99.042105} {\bibfield  {journal} {\bibinfo  {journal} {Physical Review E}\ }\textbf {\bibinfo {volume} {99}},\ \bibinfo {pages} {042105} (\bibinfo {year} {2019})}\BibitemShut {NoStop}%
\bibitem [{\citenamefont {Goldstein}\ \emph {et~al.}(2006)\citenamefont {Goldstein}, \citenamefont {Lebowitz}, \citenamefont {Tumulka},\ and\ \citenamefont {Zangh{\`\i}}}]{goldstein2006canonical}%
  \BibitemOpen
  \bibfield  {author} {\bibinfo {author} {\bibfnamefont {S.}~\bibnamefont {Goldstein}}, \bibinfo {author} {\bibfnamefont {J.~L.}\ \bibnamefont {Lebowitz}}, \bibinfo {author} {\bibfnamefont {R.}~\bibnamefont {Tumulka}}, \ and\ \bibinfo {author} {\bibfnamefont {N.}~\bibnamefont {Zangh{\`\i}}},\ }\href {\doibase 10.1103/PhysRevLett.96.050403} {\bibfield  {journal} {\bibinfo  {journal} {Physical review letters}\ }\textbf {\bibinfo {volume} {96}},\ \bibinfo {pages} {050403} (\bibinfo {year} {2006})}\BibitemShut {NoStop}%
\bibitem [{\citenamefont {Popescu}\ \emph {et~al.}(2006)\citenamefont {Popescu}, \citenamefont {Short},\ and\ \citenamefont {Winter}}]{popescu2006entanglement}%
  \BibitemOpen
  \bibfield  {author} {\bibinfo {author} {\bibfnamefont {S.}~\bibnamefont {Popescu}}, \bibinfo {author} {\bibfnamefont {A.~J.}\ \bibnamefont {Short}}, \ and\ \bibinfo {author} {\bibfnamefont {A.}~\bibnamefont {Winter}},\ }\href {\doibase 10.1038/nphys444} {\bibfield  {journal} {\bibinfo  {journal} {Nature Physics}\ }\textbf {\bibinfo {volume} {2}},\ \bibinfo {pages} {754} (\bibinfo {year} {2006})}\BibitemShut {NoStop}%
\bibitem [{\citenamefont {Calabrese}\ and\ \citenamefont {Cardy}(2005)}]{calabrese2005evolution}%
  \BibitemOpen
  \bibfield  {author} {\bibinfo {author} {\bibfnamefont {P.}~\bibnamefont {Calabrese}}\ and\ \bibinfo {author} {\bibfnamefont {J.}~\bibnamefont {Cardy}},\ }\href {\doibase 10.1088/1742-5468/2005/04/P04010} {\bibfield  {journal} {\bibinfo  {journal} {Journal of Statistical Mechanics: Theory and Experiment}\ }\textbf {\bibinfo {volume} {2005}},\ \bibinfo {pages} {P04010} (\bibinfo {year} {2005})}\BibitemShut {NoStop}%
\bibitem [{\citenamefont {Spar}\ \emph {et~al.}(2022)\citenamefont {Spar}, \citenamefont {Guardado-Sanchez}, \citenamefont {Chi}, \citenamefont {Yan},\ and\ \citenamefont {Bakr}}]{spar2022realization}%
  \BibitemOpen
  \bibfield  {author} {\bibinfo {author} {\bibfnamefont {B.~M.}\ \bibnamefont {Spar}}, \bibinfo {author} {\bibfnamefont {E.}~\bibnamefont {Guardado-Sanchez}}, \bibinfo {author} {\bibfnamefont {S.}~\bibnamefont {Chi}}, \bibinfo {author} {\bibfnamefont {Z.~Z.}\ \bibnamefont {Yan}}, \ and\ \bibinfo {author} {\bibfnamefont {W.~S.}\ \bibnamefont {Bakr}},\ }\href {\doibase 10.1103/PhysRevLett.128.223202} {\bibfield  {journal} {\bibinfo  {journal} {Physical review letters}\ }\textbf {\bibinfo {volume} {128}},\ \bibinfo {pages} {223202} (\bibinfo {year} {2022})}\BibitemShut {NoStop}%
\bibitem [{\citenamefont {Carcy}\ \emph {et~al.}(2021)\citenamefont {Carcy}, \citenamefont {Herc{\'e}}, \citenamefont {Tenart}, \citenamefont {Roscilde},\ and\ \citenamefont {Cl{\'e}ment}}]{carcy2021certifying}%
  \BibitemOpen
  \bibfield  {author} {\bibinfo {author} {\bibfnamefont {C.}~\bibnamefont {Carcy}}, \bibinfo {author} {\bibfnamefont {G.}~\bibnamefont {Herc{\'e}}}, \bibinfo {author} {\bibfnamefont {A.}~\bibnamefont {Tenart}}, \bibinfo {author} {\bibfnamefont {T.}~\bibnamefont {Roscilde}}, \ and\ \bibinfo {author} {\bibfnamefont {D.}~\bibnamefont {Cl{\'e}ment}},\ }\href {\doibase 10.1103/PhysRevLett.126.045301} {\bibfield  {journal} {\bibinfo  {journal} {Physical Review Letters}\ }\textbf {\bibinfo {volume} {126}},\ \bibinfo {pages} {045301} (\bibinfo {year} {2021})}\BibitemShut {NoStop}%
\bibitem [{\citenamefont {Bloch}\ \emph {et~al.}(2008)\citenamefont {Bloch}, \citenamefont {Dalibard},\ and\ \citenamefont {Zwerger}}]{bloch2008many}%
  \BibitemOpen
  \bibfield  {author} {\bibinfo {author} {\bibfnamefont {I.}~\bibnamefont {Bloch}}, \bibinfo {author} {\bibfnamefont {J.}~\bibnamefont {Dalibard}}, \ and\ \bibinfo {author} {\bibfnamefont {W.}~\bibnamefont {Zwerger}},\ }\href {\doibase 10.1103/RevModPhys.80.885} {\bibfield  {journal} {\bibinfo  {journal} {Reviews of modern physics}\ }\textbf {\bibinfo {volume} {80}},\ \bibinfo {pages} {885} (\bibinfo {year} {2008})}\BibitemShut {NoStop}%
\bibitem [{\citenamefont {Eisert}\ \emph {et~al.}(2008)\citenamefont {Eisert}, \citenamefont {Cramer},\ and\ \citenamefont {Plenio}}]{eisert2008area}%
  \BibitemOpen
  \bibfield  {author} {\bibinfo {author} {\bibfnamefont {J.}~\bibnamefont {Eisert}}, \bibinfo {author} {\bibfnamefont {M.}~\bibnamefont {Cramer}}, \ and\ \bibinfo {author} {\bibfnamefont {M.~B.}\ \bibnamefont {Plenio}},\ }\href@noop {} {\bibfield  {journal} {\bibinfo  {journal} {arXiv preprint arXiv:0808.3773}\ } (\bibinfo {year} {2008})}\BibitemShut {NoStop}%
\bibitem [{\citenamefont {Proctor}\ \emph {et~al.}(2022)\citenamefont {Proctor}, \citenamefont {Rudinger}, \citenamefont {Young}, \citenamefont {Nielsen},\ and\ \citenamefont {Blume-Kohout}}]{proctor2022measuring}%
  \BibitemOpen
  \bibfield  {author} {\bibinfo {author} {\bibfnamefont {T.}~\bibnamefont {Proctor}}, \bibinfo {author} {\bibfnamefont {K.}~\bibnamefont {Rudinger}}, \bibinfo {author} {\bibfnamefont {K.}~\bibnamefont {Young}}, \bibinfo {author} {\bibfnamefont {E.}~\bibnamefont {Nielsen}}, \ and\ \bibinfo {author} {\bibfnamefont {R.}~\bibnamefont {Blume-Kohout}},\ }\href {\doibase 10.1038/s41567-021-01409-7} {\bibfield  {journal} {\bibinfo  {journal} {Nature Physics}\ }\textbf {\bibinfo {volume} {18}},\ \bibinfo {pages} {75} (\bibinfo {year} {2022})}\BibitemShut {NoStop}%
\bibitem [{\citenamefont {Wallman}\ and\ \citenamefont {Emerson}(2016)}]{wallman2016noise}%
  \BibitemOpen
  \bibfield  {author} {\bibinfo {author} {\bibfnamefont {J.~J.}\ \bibnamefont {Wallman}}\ and\ \bibinfo {author} {\bibfnamefont {J.}~\bibnamefont {Emerson}},\ }\href {\doibase 10.1103/PhysRevA.94.052325} {\bibfield  {journal} {\bibinfo  {journal} {Physical Review A}\ }\textbf {\bibinfo {volume} {94}},\ \bibinfo {pages} {052325} (\bibinfo {year} {2016})}\BibitemShut {NoStop}%
\bibitem [{\citenamefont {Hayes}\ \emph {et~al.}(2020)\citenamefont {Hayes}, \citenamefont {Stack}, \citenamefont {Bjork}, \citenamefont {Potter}, \citenamefont {Baldwin},\ and\ \citenamefont {Stutz}}]{hayes2020eliminating}%
  \BibitemOpen
  \bibfield  {author} {\bibinfo {author} {\bibfnamefont {D.}~\bibnamefont {Hayes}}, \bibinfo {author} {\bibfnamefont {D.}~\bibnamefont {Stack}}, \bibinfo {author} {\bibfnamefont {B.}~\bibnamefont {Bjork}}, \bibinfo {author} {\bibfnamefont {A.}~\bibnamefont {Potter}}, \bibinfo {author} {\bibfnamefont {C.}~\bibnamefont {Baldwin}}, \ and\ \bibinfo {author} {\bibfnamefont {R.}~\bibnamefont {Stutz}},\ }\href {\doibase 10.1103/PhysRevLett.124.170501} {\bibfield  {journal} {\bibinfo  {journal} {Physical Review Letters}\ }\textbf {\bibinfo {volume} {124}},\ \bibinfo {pages} {170501} (\bibinfo {year} {2020})}\BibitemShut {NoStop}%
\end{thebibliography}

%


\appendix
\onecolumngrid

\section{Second order of entropy density evolution \label{appendix}}

We have the expansion at second order in $\delta H$
\begin{equation}
    \begin{aligned}
        e^{it(H+\delta H)}=&e^{itH}\\
        &+i\int_0^t \D{s}e^{isH}\delta H e^{i(t-s)H}\\
        &-\int_0^t \D{s} \int_0^s \D{u} e^{iuH}\delta H e^{i(s-u)H}\delta H e^{i(t-s)H}+\mathcal{O}(\delta H^3)\,.
    \end{aligned}
\end{equation}
So we get
\begin{equation}
\begin{aligned}
    e^{it(H+\delta H)} \rho e^{-it(H+\delta H)}=&e^{itH} \rho e^{-itH}\\
    &+i\int_0^t \D{s} e^{isH} [\delta H, e^{i(t-s)H}\rho e^{-i(t-s)H}]e^{-isH}\\
    &-\int_0^t \D{s} \int_0^s \D{u} e^{iuH}[\delta H, e^{i(s-u)H}[\delta H, e^{i(t-s)H}\rho e^{-i(t-s)H}]e^{-i(s-u)H} ] e^{-iuH}+\mathcal{O}(\delta H^3)\,.
\end{aligned}
\end{equation}
The entropy for a traceless perturbation is expanded as
\begin{equation}
\begin{aligned}
    S[\rho+\delta \rho]=S[\rho]-\tr_A[\log \rho_A \delta \rho_A]-\frac{1}{2}\tr_A[\delta \rho_A \rho^{-1}_A\delta \rho_A]+\mathcal{O}(\delta\rho_A^3)\,.
\end{aligned}
\end{equation}
Let us first consider the contributions coming from $\tr_A[\log \rho_A \delta \rho_A]$, with $\delta \rho_A$ of order $\delta H^2$. Using that $\rho_A\propto e^{-\beta H_A}$, and that $\tr_A[\delta\rho_A]=0$, we have
\begin{equation}
    -\tr_A[\log \rho_A \delta \rho_A]=\beta \tr[H_A \delta\rho]\,.
\end{equation}
Then, using that $\rho$ commutes with $H$, we have the second order term
\begin{equation}
    \begin{aligned}
        -\tr_A[\log \rho_A \delta \rho_A]=&...-\beta \int_0^t \D{s} \int_0^s \D{u}\tr[H_A e^{iuH}\delta H e^{i(s-u)H}\delta H\rho e^{-isH}]\\
        &+\beta \int_0^t \D{s} \int_0^s \D{u}\tr[H_A e^{iuH}\delta H e^{i(s-u)H}\rho\delta H e^{-isH}]\\
        &-\beta\int_0^t \D{s} \int_0^s \D{u} \tr[H_A e^{isH}\rho\delta H e^{-i(s-u)H}\delta H e^{-iuH}]\\
        &+\beta\int_0^t \D{s} \int_0^s \D{u} \tr[H_A e^{isH}\delta H \rho e^{-i(s-u)H}\delta H e^{-iuH}]
    \end{aligned}
\end{equation}
where $...$ denotes the zeroth and first order in $\delta H$, treated in the main text. Commuting $\rho$ and $H_A$, with an error $O(|\partial A|/N_A)$, this can be written as
\begin{equation}
    \begin{aligned}
        -\tr_A[\log \rho_A \delta \rho_A]=&...-\beta \int_0^t \D{s} \int_0^t \D{u}\tr[H_A e^{iuH}\delta H e^{i(s-u)H}\delta H \rho e^{-isH}]\\
        &+\beta \int_0^t \D{s} \int_0^t \D{u}\tr[H_A e^{iuH}\delta H e^{i(s-u)H}\rho \delta H  e^{-isH}]\,.
    \end{aligned}
\end{equation}
Let us decompose the trace into eigenstates $|E\rangle$ of $H$ with energy $E$. We have
\begin{equation}
     -\tr_A[\log \rho_A \delta \rho_A]=...+\frac{\beta}{Z} \int_0^t \D{s} \int_0^t \D{u} \sum_{E,E',E''} \langle E|H_A |E'\rangle \langle E' | \delta H |E'' \rangle \langle E''|\delta H |E\rangle e^{iu(E'-E'')}e^{is(E''-E)}(e^{-\beta E''}-e^{-\beta E})\,,
\end{equation}
with $Z$ the partition function
\begin{equation}
    Z=\sum_E e^{-\beta E}\,.
\end{equation}
Computing the integrals, we get
\begin{equation}
\begin{aligned}
     -\tr_A[\log \rho_A \delta \rho_A]=...+\frac{\beta}{Z}  \sum_{E,E',E''} &\langle E|H_A |E'\rangle \langle E' | \delta H |E'' \rangle \langle E''|\delta H |E\rangle e^{it(E'-E)/2}(e^{-\beta E''}-e^{-\beta E})\\
    & \times t^2\sinc(\tfrac{t(E'-E'')}{2})\sinc(\tfrac{t(E''-E)}{2})\,,
\end{aligned}
\end{equation}
with
\begin{equation}
    \sinc(x)=\frac{\sin x}{x}\,.
\end{equation}
The quantity $(e^{-\beta E''}-e^{-\beta E})$ behaves as $\propto (E''-E)$ for $E$ close to $E''$. Hence the quantity $(e^{-\beta E''}-e^{-\beta E})t \sinc(\tfrac{t(E''-E)}{2})$ is bounded with $t$, uniformly for $E,E''$ in a given same interval. It follows that the second order term in $\delta H^2$ in $ -\tr_A[\log \rho_A \delta \rho_A]$ is actually of order $\mathcal{O}(t \delta H^2)$, and not $\mathcal{O}(t^2 \delta H^2)$ as a priori expected.\\

Let us now focus on the term $-\frac{1}{2}\tr[\delta \rho_A \rho^{-1}_A\delta \rho_A]$ with the first order value for $\delta\rho_A$. We can write at order $\delta H^2$
\begin{equation}
     -\tr[\delta \rho_A \rho^{-1}_A\delta \rho_A]=\int_0^t \D{s}\int_0^t \D{u} \tr\left(\tr_{\bar{A}}(e^{isH}[\delta H,\rho]e^{-isH})\rho_A^{-1}\tr_{\bar{A}}(e^{iuH}[\delta H,\rho]e^{-iuH})\right)\,.
\end{equation}
Let us look at the quantities
\begin{equation}
    \int_0^t \D{s} e^{isH}[\delta H,\rho]e^{-isH}\,.
\end{equation}
Decomposed into eigenstates of $H$, this is
\begin{equation}
    \int_0^t \D{s} e^{isH}[\delta H,\rho]e^{-isH}=\sum_{E,E'} |E\rangle \langle E'| e^{it(E-E')/2} t\sinc(\tfrac{t(E-E')}{2}) \langle E| [\delta H,\rho]|E'\rangle\,.
\end{equation}
We have $\langle E| [\delta H,\rho]|E'\rangle=(e^{-\beta E'}-e^{-\beta E})\langle E|\delta H|E'\rangle/Z$. Hence
\begin{equation}
    \int_0^t \D{s} e^{isH}[\delta H,\rho]e^{-isH}=\frac{1}{Z}\sum_{E,E'} |E\rangle \langle E'| e^{it(E-E')/2} t\sinc(\tfrac{t(E-E')}{2})(e^{-\beta E'}-e^{-\beta E}) \langle E| \delta H|E'\rangle\,.
\end{equation}
Using the same argument as before, the quantity $t\sinc(\tfrac{t(E-E')}{2})(e^{-\beta E'}-e^{-\beta E})$ is bounded with $t$ uniformly in $E,E'$ in the same bounded interval. Hence we have
\begin{equation}
    -\tr[\delta \rho_A \rho^{-1}_A\delta \rho_A]=\mathcal{O}(t^0 \delta H^2)\,.
\end{equation}
In total, we thus get that the order $\delta H^2$ in the entropy density scales as $\mathcal{O}(t)$ with time, and not as $\mathcal{O}(t^2)$ as a priori expected.

\end{document}